\documentclass[11pt]{article}
\pdfoutput=1
\usepackage[centertags]{amsmath}
\usepackage[square, comma, sort&compress,numbers]{natbib}
\usepackage{array,multirow}
\numberwithin{equation}{section}
\usepackage{amssymb,amsfonts}
\usepackage{graphicx}
\usepackage{color}
\usepackage{mathtools,bm}
\usepackage{accents}
\usepackage{amsmath}
\usepackage{cancel}

\usepackage{epsfig}
\usepackage{bbold}
\usepackage{wrapfig}
\usepackage{float}
\usepackage{soul}
\usepackage{tkz-euclide}
\usepackage{braket}
\usepackage{tikz,pgf}
\usetikzlibrary{shapes}
\usetikzlibrary{calc}
\usetikzlibrary{decorations.pathmorphing}
\usetikzlibrary{decorations.pathreplacing,shapes.misc}
\usetikzlibrary{positioning}
\usetikzlibrary{arrows}
\usetikzlibrary{decorations.markings}
\usetikzlibrary{shadings}

\usetikzlibrary{intersections}

\newcommand{\be}{\begin{equation}}
\newcommand{\ee}{\end{equation}}
\newcommand{\ba}{\begin{array}}
\newcommand{\ea}{\end{array}}

\newcommand{\dps}{\displaystyle}
\newcommand{\half}{\frac{1}{2}}

\newcommand{\bref}[1]{\textbf{\ref{#1}}}

\newcommand{\im}{\mathop{\mathrm{Im}}}
\newcommand{\re}{\mathop{\mathrm{Re}}}

\newcommand{\CC}{\mathbb{C}}

\newcommand{\RR}{\mathbb{R}}

\newcommand{\cA}{\mathcal{A}}

\newcommand{\cV}{\mathcal{V}}

\numberwithin{equation}{section} \makeatletter
\@addtoreset{equation}{section}

\newcommand{\ads}{AdS$_2\;$}

\newcommand{\dm}{(\delta {m}^2)}

\newcommand{\sltwo}{sl(2,\mathbb{R})}

\newcommand{\bx}{{\bf x}}
\newcommand{\dc}[1]{\cA^{\text{cont}}_{#1}}
\newcommand{\tdc}[1]{\widetilde{\cA}^{\text{cont}}_{#1}}

\usepackage{jheppub}
\makeatletter
\def\@fpheader{\vspace{-.1cm}}
\makeatother

\title{\centering{Reduction technique for expanding the Feynman diagrams in AdS$_2$}}

\author[a,b]{Vladimir\ Khiteev}

\affiliation[a]{I.E. Tamm Department of Theoretical Physics, \\P.N. Lebedev Physical
Institute, 119991 Moscow, Russia}
\affiliation[b]{Moscow Institute of Physics and Technology, \\
Institutsky lane 9, 141700, Dolgoprudny, Moscow region, Russia}

\emailAdd{khiteev.vs@phystech.edu}

\abstract{We formulate the reduction technique for expanding the $n$-point AdS Feynman diagrams in two dimensions into the Wilson line networks. The technique applies to any $n$-point tree diagram, expressing it as several series of the matrix elements of Wilson line network operators, and employs the Wilson network expansions of the $(n-1)$-point AdS$_2$ Feynman diagrams, which can in turn be obtained using the same technique. Consequently, the complexity of this technique does not increase with $n$, provided that the expansions of the $(n-1)$-point diagrams are known. To demonstrate the reduction technique, we apply it to all topologically distinct five-point AdS$_2$ Feynman tree diagrams. The resulting expansions near the conformal boundary reproduce the known decompositions of the corresponding five-point Witten diagrams into conformal blocks. }

\begin{document}

\maketitle
\flushbottom

\section{Introduction}

The question of recovering bulk dynamics of anti-de Sitter (AdS) space from the boundary conformal field theory (CFT) has been considered since the early days of AdS/CFT \cite{Witten:1998qj, Balasubramanian:1998sn, Balasubramanian:1998de, Banks:1998dd, Hamilton:2005ju, Hamilton:2006az, Czech:2016xec}. Most of these works study the reconstruction of the bulk fields, while the question of the reconstruction of the AdS correlators is still mostly unanswered. Hamilton et al. \cite{Hamilton:2005ju, Hamilton:2006az} suggested a way to reconstruct the bulk correlation functions, but their analysis covers only correlators of one bulk field in the $\lambda\phi^3$ theory and several boundary fields. To avoid considering a particular theory, one can work with the AdS Feynman diagrams and reconstruct them from the conformal boundary.

In this paper, we continue our study of AdS$_2$ Feynman tree diagrams in the context of the AdS$_2$/CFT$_1$ correspondence \cite{Alkalaev:2026gkq,Alkalaev:2025meb, Alkalaev:2024cje, Alkalaev:2023axo}. The restriction to the tree-level is motivated by the $1/N$ expansion.\footnote{From the boundary CFT perspective, the role of $N$ is played by the central charge $c$. In the limit $c\to\infty$ the Virasoro algebra contracts to  $\sltwo$ subalgebra and the Virasoro conformal blocks are reduced to global conformal blocks provided all operators are light (i.e. their conformal weights grow as $c^0$). Considering loop diagrams corresponds to including  corrections in powers of $1/c$ \cite{Besken:2017fsj}.} Specifically, we decompose the AdS Feynman tree diagrams in two dimensions into the matrix elements of Wilson networks, which we refer to as AdS$_2$ vertex functions. These functions are holographically reconstructed global conformal blocks. The purpose of such a decomposition is twofold. Firstly, since the AdS$_2$ Feynman propagator is no longer a simple rational function as in the Minkowski case but is given by the Gauss hypergeometric function, the calculation of the Feynman diagrams in the AdS$_2$ spacetime remains a difficult problem. Such a decomposition circumvents the AdS$_2$ integration and provides an alternative representation of AdS$_2$ Feynman diagrams. Secondly, this decomposition is important for understanding the AdS/CFT correspondence, since it provides a constructive way to reconstruct bulk observables (AdS$_2$ Feynman diagrams) from the boundary ones (global conformal blocks) and vice versa. Notably, conformal block decompositions of the Witten diagrams can be obtained from the Wilson network expansions of AdS$_2$ Feynman diagrams by considering the boundary asymptotics of the latter.

Three-point and four-point AdS$_2$ Feynman tree diagrams can be expanded into the Wilson networks using the decomposition method formulated in \cite{Alkalaev:2026gkq}. It consists of two parts: (1) applying specific propagator identities to every external leg of an AdS$_2$ Feynman diagram, producing a finite sum of modified AdS$_2$ Feynman diagrams composed of standard bulk-to-bulk propagators and modified propagators, and (2) converting the resulting modified diagrams into infinite sums of the AdS$_2$ vertex functions using the integral representations of the latter. There are two problems with applying this method to higher-point  diagrams. The first one is the large number of intermediate terms produced by part (1) of the algorithm, which grows exponentially with $n$. This complicates the decomposition procedure since each intermediate term must be treated separately.

Another problem lies in the structure of the algorithm itself. The propagator identities are applied only to the external bulk-to-bulk propagators, i.e. those connecting an endpoint of the diagram to its internal vertex. Thus, if the diagram has many external bulk-to-bulk propagators, a large portion of the diagram is left untouched by the first part of the decomposition algorithm. This is not a problem in itself, but part (2) of the algorithm requires converting the modified diagrams produced by part (1) into the AdS$_2$ vertex functions using the integral representations of the latter. Since modified AdS$_2$ Feynman diagrams with an unmodified inner part do not admit simple representations in terms of AdS$_2$ vertex functions and should be decomposed into infinite sums of the latter, the application of part (2) of the algorithm is hindered. This is not an issue in the three-point and four-point cases, since the number of internal propagators (i.e. those connecting two internal vertices of the diagram) is at most one.

To circumvent these problems, we formulate the reduction technique for expanding $n$-point AdS$_2$ Feynman diagrams. It is based on the decomposition method used previously and consists of three parts: (I) applying part (1) of the previous method to two external legs of an $n$-point AdS$_2$ Feynman diagram sharing a common vertex, which results in a sum of integrals of $(n-1)$-point AdS$_2$ Feynman diagrams; (II) substituting the Wilson network expansions of these $(n-1)$-point diagrams, resulting in a sum of integrals of $(n-1)$-point AdS$_2$ vertex functions; (III) applying the integral relations between $(n-1)$-point and $n$-point AdS$_2$ vertex functions to convert the former into the latter. This technique is recursive: it employs the Wilson network expansions of $(n-1)$-point AdS$_2$ Feynman diagrams that can be obtained in the same way until one reaches the four-point case, which is already solved. 

The reduction technique resolves the first problem, which is the large number of intermediate terms, by limiting the number of propagators on which we act to two. This ensures that the number of terms produced by part (1) of the decomposition method is bounded, independently of the number of external propagators in the diagram. The second problem, expanding the modified AdS$_2$ Feynman diagrams into AdS$_2$ vertex functions, is reduced to finding integral relations between $(n-1)$-point and $n$-point AdS$_2$ vertex functions, which involve only one-dimensional integrals or contour integrals over finite domains. We provide a complete set of these integral relations for the $n$-point AdS$_2$ vertex functions in the comb channel.

To demonstrate the reduction technique, we decompose all five-point AdS$_2$ Feynman tree diagrams. Since the resulting expressions become unwieldy, we formulate an algorithm for obtaining the terms of the expansion. This algorithm, derived from the analysis of expansions of three-, four- and five-point AdS$_2$ Feynman diagrams, allows one to read off a particular expansion term from the original diagram.

Note that the restriction to two dimensions is motivated by the technical simplicity: AdS$_2$ provides the simplest setting for studying Wilson network expansions, and we take it as the starting point for developing the method of finding the Wilson network expansions of AdS$_2$ Feynman diagrams before addressing the more involved higher-dimensional cases. To generalize the procedure for the Wilson network expansion one requires generalizing two elements of the construction. The first one is to define and calculate the Wilson line networks in $d$ dimensions. The second one is to formulate the propagator identities in $d$ dimensions. Although some propagator identities can be readily formulated in higher dimensions, generalizing the other identities is not straightforward. We leave both problems for future work. 

The paper is organized as follows. In Section \bref{sec:overview}, we review the expansion of the four-point contact AdS$_2$ Feynman diagram, establish our notation and also provide the decomposition method from \cite{Alkalaev:2026gkq} used to expand three- and four-point diagrams. In Section \bref{sec:reduction}, we formulate the reduction technique for expanding the $n$-point AdS$_2$ Feynman diagrams. In Section \bref{sec:decomp}, we apply the reduction technique to expand all five-point AdS$_2$ Feynman tree diagrams and study the resulting decompositions. Subsections \bref{sec:5_cont}, \bref{sec:5_ex} present new Wilson network expansions of the three five-point AdS$_2$ Feynman diagrams. Due to the large number of terms in these expansions, certain coefficients and the arrangements of the conformal weights of the AdS$_2$ vertex functions are listed in Appendix \bref{app:coefs}. In subsection \bref{sec:rules}, we formulate an algorithm for obtaining the terms in the decomposition of various AdS$_2$ Feynman diagrams. In subsection \bref{sec:boundary}, we compare the boundary asymptotics of the resulting expansions with the known conformal block decompositions of five-point Witten diagrams. The application of the reduction technique to the decomposition of the $n$-point AdS$_2$ Feynman diagrams is briefly discussed in Section \bref{sec:conclusion}. A detailed derivation of integral relations between $n$-point and $(n-1)$-point AdS$_2$ vertex functions is given in Appendix \bref{app:derivations}.

\section{Decomposition of AdS$_2$  diagrams}
\label{sec:overview}

\subsection{Wilson network expansion of four-point contact AdS$_2$  Feynman diagram}

Bulk correlation functions of AdS fields can be decomposed into AdS Feynman diagrams which involve multiple integrals over AdS space with integrands given by products of bulk-to-bulk propagators. As was shown in \cite{Alkalaev:2026gkq, Alkalaev:2025meb}, three- and four-point diagrams can be decomposed into matrix elements of the Wilson line networks, which can be identified with HKLL-reconstructed global conformal blocks.\footnote{HKLL construction provides a way to reconstruct scalar fields in the bulk from the primary fields located at the boundary by integrating them with the smearing functions \eqref{smear} \cite{Hamilton:2005ju, Hamilton:2006az}. Here, we use a similar construction, where we reconstruct AdS$_2$  vertex functions in the bulk from the conformal blocks on the boundary, see \cite{Alkalaev:2024cje} for details.} The matrix elements of the Wilson line networks can be viewed as a convenient basis in the space of bulk correlation functions, enabling a systematic decomposition of AdS$_2$ Feynman diagrams, just as conformal blocks provide a basis for decomposing conformal correlation functions in the boundary CFT$_1$. For example, the decomposition of the four-point contact AdS$_2$  Feynman diagram is given by
\be 
\label{4pt_cont_decomp}
\ba{l}
\dps
\dc{h_1...h_4}(\bx_i) = \int_{\text{AdS}_2}d^2\bx\;\sqrt{g(\bx)}\; \prod_{i=1}^4 G_{h_i}(\bx,\bx_i) = 
\sum_{n=0}^{\infty} a^{h_1h_2}_n\gamma_{h_{12|n}h_3h_4}\, 
\cV_{h_1h_2 h_3 h_4,h_{12|n}}(\bx_i)
\vspace{3mm}
\\
\dps
\hspace{17mm}
+\sum_{n=0}^{\infty} \gamma_{h_1h_2h_{34|n}}a^{h_3h_4}_n\,
\cV_{h_1h_2 h_3 h_4,h_{34|n}}(\bx_i)
\vspace{3mm}
\\
\dps
\hspace{17mm}
+\sum_{n,m=0}^{\infty}
\frac{a^{h_2h_{34|n}}_ma^{h_3h_4}_n}{\dm^{h_{234|nm}}_{h_1}}\,
\cV_{h_{234|nm} h_2 h_3 h_4,h_{34|n}}(\bx_i) + (1\leftrightarrow 2) 
\vspace{3mm}
\\
\dps
\hspace{17mm}
+\sum_{n,m=0}^{\infty}
\frac{a^{h_1h_2}_na^{h_4h_{12|n}}_m}{\dm^{h_{124|nm}}_{h_3}}\,
\cV_{h_1h_2 h_{124|nm} h_4,h_{12|n}}(\bx_i) + (3\leftrightarrow 4)\,.
\ea
\ee
The following elements appear in the expansion above:

\noindent {\bf 1.} The metric of the Euclidean AdS$_2$ in Poincaré coordinates $\bx = (u,z)$ is given by 
\be 
\label{metric}
ds^2 = \frac{dz^2+du^2}{u^2}\,,
\qquad
u\in\mathbb{R}_{\geq 0}\,,\; z \in \mathbb{R}\,.
\ee
The conformal boundary of global \ads is disconnected; we choose the component defined by $u=0$. The metric determinant is $g(\bx) = u^{-4}$. From now on, we omit the subscript denoting the AdS dimension.

\noindent {\bf 2.} $G_h(\bx,\bx')$ is the bulk-to-bulk propagator of a scalar field of mass $m^2 = h(h-1)$, $h\geq\half$ \cite{Fronsdal:1974ew}:
\be 
\label{bulk-to-bulk}
G_h(\bx,\bx') = \left(\frac{\xi(\bx,\bx')}{2}\right)^h{}_2F_1\left[\frac{h}{2},\frac{h}{2}+\half; h+\half\Big|\,\xi(\bx,\bx')^2\right],\quad \xi(\bx,\bx')= \frac{2u u'}{u^2+u'^2+(z-z')^2}\,.
\ee

\noindent {\bf 3.} The functions $\cV_{h_1h_2 h_3 h_4,\tilde{h}}(\bx_i)$ are four-point AdS vertex functions, which are the matrix elements of four-point Wilson line networks. Such a network carries four external weights $h_1,...,h_4$ and a single internal weight $\tilde{h}$. The construction of the general $n$-point Wilson line network is thoroughly discussed in \cite{Alkalaev:2023axo} (see also \cite{Bhatta:2016hpz,Besken:2016ooo,Bhatta:2018gjb,Castro:2018srf,Alkalaev:2020yvq}). Here we only note that the AdS vertex functions admit an explicit representation as multivariate generalized hypergeometric series \cite{Alkalaev:2024cje}.\footnote{Here we use a rescaled version of the AdS vertex functions obtained by dividing the expression in \cite{Alkalaev:2024cje} by the coefficient $C_{h,\tilde{h}}$, the explicit form of which is given by eq. $(2.18)$ in \cite{Alkalaev:2024cje}.}

\noindent {\bf 4.} The conformal weights are subject to the triangle inequalities 
\be
\label{triangle_identity_4pt}
h_1 + \dots - h_k + \dots + h_4 > 0\,, 
\qquad 
\forall k \in \{1, ...\,, 4\}\,.
\ee
By a slight abuse of notation, we collectively denote the external weights $h_1, ..., h_4$ as $h$. We also define the ``multi-trace conformal weight'' as $h_{i_1...i_k|n_1...n_{k-1}} = h_{i_1}+...+h_{i_k}+2n_1+...+2n_{k-1}$. 

\noindent {\bf 5.} The coefficients are given by
\be
\label{a_beta}
\ba{c}
\dps
a^{h_1h_2}_m = \frac{(-)^m(h_1)_m(h_2)_m}{m!(h_1+h_2-\half+m)_m}\,,
\quad
\dm^{h_1}_{h_2} = \frac{\Gamma(h_2)(h_1(h_1-1) - h_2(h_2-1))}{2\pi^{\half}\Gamma(h_2+\half)}\;,
\vspace{3mm}
\\
\dps
\gamma_{h_1h_2h_3} = \frac{\pi^\half}{2}\Gamma\Big(\frac{h_1+h_2+h_3}{2}-\half\Big)\frac{\Gamma(\frac{h_1+h_2-h_3}{2})\Gamma(\frac{h_1-h_2+h_3}{2})\Gamma(\frac{-h_1+h_2+h_3}{2})}{\Gamma(h_1)\Gamma(h_2)\Gamma(h_3)}\,,
\ea
\ee
and $(a)_n = \Gamma(a+n)/\Gamma(a)$ is the Pochhammer symbol. By definition, $\dm^{h_1}_{h_2}$ is proportional to the difference between two mass terms corresponding to the quadratic Casimir operator eigenvalues.

\subsection{Decomposition algorithm}
\label{sec:dec_alg}

The expansion \eqref{4pt_cont_decomp} along with similar expansions of the three-point contact and the four-point exchange AdS Feynman diagrams can be obtained by applying the decomposition algorithm formulated in \cite{Alkalaev:2026gkq}. It consists of two parts: (i) applying specific propagator identities to the bulk-to-bulk propagators inside the Feynman diagram, producing a sum of modified AdS Feynman diagrams, and (ii) converting these diagrams into AdS vertex functions using integral representations of the latter. We briefly review the decomposition algorithm below.

\paragraph{AdS propagator integral identities.}  The complete set of identities required for obtaining \eqref{4pt_cont_decomp} consists of five relations. Propagators are distributions, since they arise as solutions to differential equations with distributional sources. Therefore, every propagator identity should be understood as an integral identity, where each side is integrated against a test function. For brevity, we omit the integrals and the test functions in the relations below wherever possible.
\begin{figure}
\centering
\includegraphics[scale=0.9]{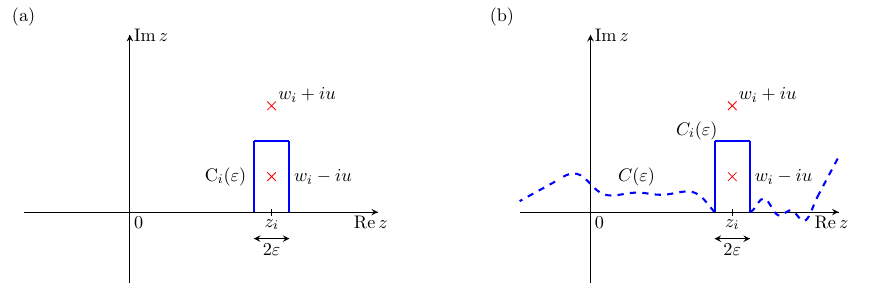}
\caption{{\bf (a)} The contours $C_i(\varepsilon)$ in the complex $z$-plane, $i=1,...,4$. The red crosses denote the poles of the modified propagators $\widehat{G}(\bx,\bx_i,w_i)$. {\bf (b)} An example of the contour $C(\varepsilon)$ (dashed line) in the complex $z$-plane. Note that the total  contour $C(\varepsilon) + C_i(\varepsilon)$ is always open due to the constraint on the contour $C(\varepsilon)$, namely $\re(z)\neq z_i$ for all $z\in C(\varepsilon)$. The contour $C(\varepsilon)$ does not cross the poles and branch cuts of the test function $f(\bx)$.}
\label{fig:C_i}
\end{figure}

\begin{figure}
\centering
\includegraphics[scale=0.9]{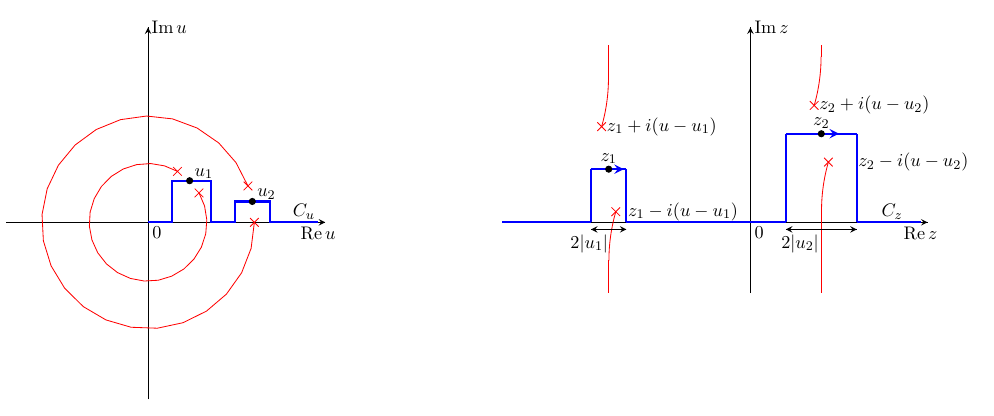}
\caption{Integration contours $C_u$ and $C_z$ in the complex $u$- and $z$-planes, respectively. The red crosses and lines denote the branch points and cuts of the integrand in \eqref{transition_id}.}
\label{fig:C_u}
\end{figure}

\begin{itemize}
\item The conversion identity: 
\be 
\label{conversion_id}
G_h(\bx,\bx') = \int_{z'-iu'}^{z'+iu'}d w\;\widehat{G}_h(\bx,\bx',w)\,,
\qquad
\re(z)\neq \re(z')\;\; \text{or} \;\; u>u'-|\im(z-z')|\,,
\ee 
where $(u,z), (u',z')\in\RR_{\geq0}\times\CC$ and the modified propagator $\widehat{G}_h(\bx,\bx',w)$ is defined by 
\be 
\label{G_hat}
\widehat{G}_h(\bx,\bx',w) = K_h(\bx, w)\mathbb{K}_{h}(\bx', w)\,,
\ee 
where the bulk-to-boundary propagator $K_h(\bx, w)$ is obtained from the bulk-to-bulk propagator \eqref{bulk-to-bulk} by taking one of the bulk points to the conformal boundary,  
\be 
\label{bulk-to-boundary}
K_h(\bx, w) = \left(\frac{u}{u^2 + (z-w)^2}\right)^h,
\ee
and the smearing function $\mathbb{K}_{h}(\bx', w)$ is proportional to the bulk-to-boundary propagator of dual weight $1-h$
\be
\label{smear}
\mathbb{K}_h(\bx',w) = \frac{-2i}{4^{h}}\frac{\Gamma(2h)}{\Gamma(h)\Gamma(h)}\left(\frac{u'}{u'^2 + (z'-w)^2}\right)^{1-h}\,.
\ee 
\item The superposition  identity:
\be 
\label{superposition_id}
\ba{l}
\dps
\lim_{\varepsilon\to0}\int_{0}^{\infty} \frac{du}{u^2}\int_{C(\varepsilon)}dz\;G_{h_i}(\bx,\bx_i)f(\bx) = \pi\int_{0}^{u_i} \frac{du}{u^2}\int_{z_i-i(u-u_i)}^{z_i+i(u-u_i)}dz\;\widetilde{G}_{h_i}(\bx,\bx_i)f(\bx)
\vspace{2.5mm}  
\\
\dps
\hspace{32mm}+\lim_{\varepsilon\to0}\int_{z_i-iu_i}^{z_i+iu_i}dw_i\int_{0}^{\infty} \frac{du}{u^2}\int_{C(\varepsilon)+C_i(\varepsilon)}dz\;\widehat{G}_{h_i}(\bx,\bx_i,w_i)f(\bx)\,,
\ea
\ee 
where the contour $C_i(\varepsilon)$ is shown in fig.~\bref{fig:C_i} {\bf (a)}; $C(\varepsilon)$ is a contour on the complex $z$-plane such that $C(\varepsilon)+C_i(\varepsilon)$ is a continuous curve; $C(\varepsilon)$ satisfies $\re(z)\neq z_i$ for all $z\in C(\varepsilon)$, see fig.~\bref{fig:C_i} {\bf (b)} for the example of such contour; $f(\bx)$ is a test function holomorphic on a domain $U \times V \subset \CC^2$, such that $ \RR_+ \subset U$ and $ C(\varepsilon)+C_i(\varepsilon) \subset V $; the modified propagator $\widetilde{G}_h(\bx,\bx')$ is given by
\be 
\label{G_tilde}
\widetilde{G}_h(\bx,\bx') = \frac{-2i}{4^{h}}\frac{\Gamma(2h)}{\Gamma(h)\Gamma(h)}\left(\frac{2^{2h}\Gamma(1-2h)}{\Gamma(1-h)^2}\, G_h(\bx,\bx') + \frac{2^{2-2h}\Gamma(2h-1)}{\Gamma(h)^2}\, G_{1-h}(\bx,\bx')\right)\,.
\ee
\item The splitting identity:
\be 
\label{geodesic_split}
\ba{l}
\dps
\pi \int_{0}^{u_3} \frac{du}{u^2} \int_{z_3-i(u-u_3)}^{z_3+i(u-u_3)}\hspace{-2mm}dz\; G_{h_1}(\bx,\bx_1) G_{h_2}(\bx,\bx_2) \widetilde{G}_{h_3}(\bx,\bx_3) = \sum_{n=0}^{\infty}\frac{a^{h_1h_2}_n\,\alpha_{h_1h_2,n}}{\dm^{h_{12|n}}_{h_3}}
\vspace{2.5mm}  
\\
\dps
\times\int_{z_3-iu_3}^{z_3+iu_3} \hspace{-2mm} dw\;\oint_{0} \frac{du}{u^2} \oint_{P[w-iu,w+iu]} \hspace{-2mm} dz\, 
G_{h_1}(\bx,\bx_1)G_{h_2}(\bx,\bx_2) \widehat{G}_{h_{12|n}}(\bx,\bx_3,w)
\vspace{3mm}
\\
\dps
= \sum_{n=0}^{\infty}  \frac{a^{h_1h_2}_n}{\dm^{h_{12|n}}_{h_3}} \cV_{h_1h_2 h_{12|n} }(\bx_1,\bx_2,\bx_3)\,,
\ea
\ee
where $P[w-iu,w+iu]$ is the Pochhammer contour and
\be 
\label{alpha_prime}
\alpha_{h_1h_2,n} = \frac{(-)^{h_1+h_2+n}}{4\pi^{\frac{3}{2}}\sin(2\pi (h_1+h_2))}\frac{n!\,\Gamma(h_{12|n})}{\Gamma(h_1+h_2+n-\half)(h_1)_n(h_2)_n}\,.
\ee 
Note that any bulk-to-bulk propagator $G_{h_i}(\bx,\bx_i)$ in the first and  second lines of \eqref{geodesic_split} can be converted to the integral of the modified propagator $\widehat{G}_{h_i}(\bx,\bx_i,w_i)$ using the conversion identity \eqref{conversion_id}.

\item The geodesic decomposition identity:\footnote{This is the bulk version of the identity from \cite{Hijano:2015zsa}.}
\be 
\label{geodesic_prop}
\ba{l}
\dps
\prod_{j=1}^2\int_{z_j-iu_j}^{z_j+iu_j}dw_j\; \widehat{G}_{h_1}(\bx,\bx_1,w_1)\widehat{G}_{h_2}(\bx,\bx_2,w_2) = \sum_{n=0}^\infty \frac{a^{h_1h_2}_n}{\beta_{h_{12|n}h_1h_2}} 
\vspace{3mm}
\\
\dps
\times \prod_{j=1}^2\int_{z_j-iu_j}^{z_j+iu_j}dw_j\;\int_{\gamma_{12}}d\lambda\, \widehat{G}_{h_1}(\bx(\lambda),\bx_1,w_1)\widehat{G}_{h_2}(\bx(\lambda),\bx_2,w_2)G_{h_{12|n}}(\bx(\lambda),\bx)\,,
\ea
\ee
where $|\xi(\bx,\bx(\lambda))|<1$ for all  $\lambda \in \RR$, $\xi(\bx,\bx(\lambda))$  is a geodesic distance \eqref{bulk-to-bulk}, the analytically continued geodesic $\bx(\lambda)\in\gamma_{12}$, $\bx(\lambda) = (u(\lambda),z(\lambda))$ connecting two points $w_1\in\CC$ and $w_2\in\CC$ is given by 
\be 
\label{comp_geod}
u(\lambda)^2 = \frac{(w_1-w_2)^2}{4\cosh^2\lambda}\,,
\qquad
z(\lambda) = \frac{(w_1+w_2)}{2} + \frac{(w_1-w_2)}{2}\tanh\lambda\,;
\ee 
and
\be 
\label{beta_coef}
\beta_{h_1h_2h_3} = \frac{\Gamma(\frac{h_1+h_2-h_3}{2})\Gamma(\frac{h_1+h_3-h_2}{2})}{2\Gamma(h_1)}\,.
\ee 
\item The transition identity:
\be 
\label{transition_id}
\int_{C_u}\frac{du}{u^2}\int_{C_z}dz\; G_{h_1}(\bx,\bx_1)G_{h_2}(\bx,\bx_2)= \frac{1}{{\dm^{h_1}_{h_2}}}G_{h_1}(\bx_1,\bx_2)+\frac{1}{{\dm^{h_2}_{h_1}}}G_{h_2}(\bx_1,\bx_2)\,,
\ee
where $\bx_1,\bx_2\in\CC\times\CC$ and $|z_1-z_2|\geq|u_1+u_2|$; the integration contours $C_u$ and $C_z$ are shown in fig.~\bref{fig:C_u}.

\end{itemize}

The first part of the decomposition algorithm consists of three steps: (i) applying the superposition identity \eqref{superposition_id} to every external bulk-to-bulk propagator of an AdS Feynman diagram and converting bulk-to-bulk propagators to the modified propagators $\widehat{G}$ using the conversion identity \eqref{conversion_id} in cases where the superposition identity is not applicable; (ii) applying the geodesic decomposition identity \eqref{geodesic_prop} to every product of two modified propagators of the form $\widehat{G}_{h_i}(\bx,\bx_i,w_i)\widehat{G}_{h_j}(\bx,\bx_j,w_j)$; (iii) applying the transition identity \eqref{transition_id} and the splitting identity \eqref{geodesic_split} wherever possible. 

\paragraph{Integral representations of the four-point AdS vertex function.} Applying the first part of the decomposition algorithm to an AdS Feynman diagram yields a sum of modified AdS Feynman diagrams. The second part of the decomposition algorithm converts these modified diagrams into the AdS vertex functions using the corresponding integral representations of the latter. The specific set of the integral representations used in a given decomposition depends on the AdS Feynman diagram. For example, to decompose the four-point contact AdS Feynman diagram one employs three types of integral representations of the four-point AdS vertex function $\cV_{h_1 h_2 h_3 h_4,\tilde{h}}(\bx_i)$. Here and everywhere below we impose the restriction $h_i,\tilde{h}\geq \half$, which is necessary for the representations to hold.

The first representation expresses the four-point AdS vertex function in terms of the three-point one
\be 
\label{alt_geodesic_rep}
\ba{c}
\dps
\cV_{h_1 h_2 h_3 h_4,\tilde{h}}(\bx_1,\bx_2,\bx_3,\bx_4) = \frac{1}{\beta_{\tilde{h} h_3 h_4}}\prod_{j=3}^4\int_{z_j-iu_j}^{z_j+iu_j}dw_j \int_{\gamma_{34}}d\lambda\, \cV_{h_1 h_2 \tilde{h}}(\bx_1,\bx_2, \bx(\lambda))
\vspace{3mm}
\\
\dps
\times\, \widehat{G}_{h_3}(\bx(\lambda),\bx_3,w_3)\,\widehat{G}_{h_4}(\bx(\lambda),\bx_4,w_4)\,,\qquad  h_3 < \tilde{h}+h_4,\quad  h_4<\tilde{h}+h_3\,,
\ea
\ee 
where $\beta_{\tilde{h} h_3 h_4}$ is given by \eqref{beta_coef}, $\cV_{h_1 h_2 \tilde{h}}(\bx_1,\bx_2, \bx(\lambda))$ is the three-point AdS vertex function, the geodesic $\bx(\lambda)\in\gamma_{34}$, $\bx(\lambda) = (u(\lambda),z(\lambda))$ connects the boundary points $z_3$ and $z_4$ \eqref{comp_geod} and the restriction on the conformal weights ensures the convergence of the integral.

The three-point AdS vertex function from the previous relation can be represented as a geodesic integral, yielding
\be 
\label{geodesic_rep}
\ba{c}
\dps
\cV_{h_1 h_2 h_3 h_4,\tilde{h}}(\bx_1,\bx_2,\bx_3,\bx_4) = \frac{1}{\beta_{\tilde{h} h_1 h_2}\beta_{\tilde{h} h_3 h_4}}\prod_{j=1}^4\int_{z_j-iu_j}^{z_j+iu_j}dw_j \int_{\gamma_{12}}d\lambda\;\int_{\gamma_{34}}d\lambda'\, \widehat{G}_{h_1}(\bx(\lambda),\bx_1,w_1)
\vspace{3mm}
\\
\dps
\times\, \widehat{G}_{h_2}(\bx(\lambda),\bx_2,w_2)\,G_{\tilde{h}}(\bx(\lambda),\bx(\lambda'))\,\widehat{G}_{h_3}(\bx(\lambda'),\bx_3,w_3)\,\widehat{G}_{h_4}(\bx(\lambda'),\bx_4,w_4)\,,
\ea
\ee 
where the weights satisfy
\be 
h_1 < \tilde{h}+h_2,\,\quad h_2<\tilde{h}+h_1,\,\quad  h_3 < \tilde{h}+h_4,\,\quad  h_4<\tilde{h}+h_3\,.
\ee 

The third integral representation has the same integrand as the first one \eqref{alt_geodesic_rep}, but the integration domain and the restrictions on the weights differ
\be 
\label{non_triangle_rep}
\ba{c}
\dps 
\cV_{h_1 h_2 h_3 h_4,\tilde{h}}(\bx_1,\bx_2,\bx_3,\bx_4) = \alpha_{\tilde{h}h_3,n} \prod_{j=3}^4\int_{z_j-iu_j}^{z_j+iu_j}dw_j \; \oint_{0} \frac{du}{u^2} \oint_{P[w_4-iu,w_4+iu]} \hspace{-2mm} dz
\vspace{3mm}
\\
\dps
\hspace{15mm}\times\,\cV_{h_1 h_2 \tilde{h}}(\bx_1,\bx_2, \bx) \,\widehat{G}_{h_3}(\bx,\bx_3,w_3)\,\widehat{G}_{h_4}(\bx,\bx_4,w_4)\,,\qquad h_4 = \tilde{h}+h_3+2n\,,
\ea
\ee
where the $\alpha$-coefficient is given by \eqref{alpha_prime}.

\paragraph{Obstructions to the $n$-point generalization of the decomposition algorithm.} The decomposition algorithm described above is formulated for the three-point and four-point AdS Feynman diagrams. Nevertheless, attempting to decompose the $n$-point AdS Feynman diagrams for $n\geq 5$ using this algorithm, we find that there are two technical difficulties obstructing the decomposition. The first one is the large number of terms produced by step (i) of the algorithm, which involves applying the superposition identity \eqref{superposition_id} to each external bulk-to-bulk propagator. Each application of this identity produces two terms, each of which can contain a bulk-to-bulk propagator suitable for applying the superposition identity again. In the result, the total number of the intermediate terms grows exponentially with $n$.\footnote{The highest number of terms produced by step (i) of the algorithm for a fixed $n$ is $9\cdot2^{n-4}$, and it is achieved for the $n$-point AdS Feynman diagram in the comb channel.} Since each intermediate term must be treated separately, this complicates the remaining steps of the decomposition procedure.

The second issue is related to the integral representations of the AdS vertex functions. Step (i) of the decomposition algorithm involves operations on the external bulk-to-bulk propagators, while internal propagators are left untouched. If the original diagram has many internal bulk-to-bulk propagators, a large portion of the diagram is unaffected by the decomposition algorithm. Such diagrams do not admit simple representations in terms of AdS vertex functions and should be decomposed into infinite sums of the latter. This requires a new method of decomposition which either involves operations on the internal bulk-to-bulk propagators or converts the internal part of the diagram to AdS vertex functions. These two obstructions motivate the reduction technique, which we formulate in the next section.

\section{Reduction technique}
\label{sec:reduction}

To circumvent the problems described above and simplify the process of decomposing the $n$-point AdS Feynman diagrams, we propose a new algorithm, which we call the reduction technique. It is based on the decomposition algorithm described above and allows obtaining the Wilson network decomposition of $n$-point diagrams using integral relations between AdS vertex functions (similar to \eqref{alt_geodesic_rep}) and the Wilson network decompositions of $(n-1)$-point diagrams. It consists of three parts: (I) applying step (i) of the decomposition algorithm described in subsection \bref{sec:dec_alg} to two external bulk-to-bulk propagators, (II) representing the result as a sum of integrals of the $(n-1)$-point (modified) AdS Feynman diagrams and replacing them with their Wilson network decompositions, (III) applying the integral relation between the $(n-1)$-point and $n$-point AdS vertex functions similar to \eqref{alt_geodesic_rep}. We demonstrate how this algorithm avoids the issues of the previous algorithm on the example of an $n$-point AdS Feynman diagram.

\paragraph{I.} Consider an arbitrary $n$-point AdS Feynman tree diagram with $n\geq 3$. It can always be represented as an integral of the lower-point diagram multiplied by bulk-to-bulk propagators:
\be
\label{general_struct}
\cA^{n,k}_{h,\tilde{h}}(\bx_i) = \int_{\text{AdS}_2}d^2\bx\;\sqrt{g(\bx)}\; G_{h_1}(\bx_1,\bx)...G_{h_k}(\bx_k,\bx)\cA^{n-k+1}_{\tilde{h}_1h_{k+1}...h_n,\tilde{h}_2...\tilde{h}_m}(\bx,\bx_{k+1},...,\bx_n)\,,
\ee 
where $k\geq 2$ is the number of the bulk-to-bulk propagators involved in the integration over $\bx$, $m$ is the number of internal bulk-to-bulk propagators, $\cA^{n-k+1}_{\tilde{h}_1h_{k+1}...h_n,\tilde{h}_2...\tilde{h}_m}(\bx,\bx_{k+1},...,\bx_n)$ is the $(n-k+1)$-point AdS Feynman tree diagram with the topology determined by the original diagram, the conformal weights $\{h_1,h_2,...,h_n\}$ and $\{\tilde{h}_1,...,\tilde{h}_m\}$ are collectively denoted by $h$ and $\tilde{h}$, respectively. Depending on the value of $k$, there are two cases for the first step of the reduction technique: the cases $k=2$ and $k\geq3$, which differ in the structure of the resulting expansion.

In the case $k=2$, one sequentially applies steps (i), (ii) and (iii) of the first part of the decomposition algorithm described in subsection \bref{sec:dec_alg} to the product of bulk-to-bulk propagators $G_{h_1}(\bx_1,\bx)G_{h_2}(\bx_2,\bx)$ in \eqref{general_struct} and obtains
$$
\ba{l}
\dps
\cA^{n,2}_{h,\tilde{h}}(\bx_i) = \frac{\gamma_{\tilde{h}_1h_1h_2}}{\beta_{\tilde{h}_1h_1h_2}}\prod_{j=1}^2\int_{z_j-iu_j}^{z_j+iu_j}dw_j \int_{\gamma_{12}}d\lambda\; \widehat{G}_{h_1}(\bx(\lambda),\bx_1,w_1)\widehat{G}_{h_2}(\bx(\lambda), \bx_2, w_2)
\vspace{3mm}
\\
\dps
\times\cA^{n-1}_{\tilde{h}_1h_3...h_n,\tilde{h}_2...\tilde{h}_m}(\bx(\lambda),\bx_3,...,\bx_n) + \lim_{\varepsilon \to 0^+}\sum_{n=0}^\infty\frac{a^{h_1h_2}_n(1-\varepsilon)^n}{\beta_{h_{12|n}h_1h_2}\dm^{h_{12|n}}_{\tilde{h}_1}}\prod_{j=1}^2\int_{z_j-iu_j}^{z_j+iu_j}dw_j 
\ea 
$$
\be
\label{reduction_ex}
\ba{l}
\dps
\times \int_{\gamma_{12}}d\lambda\; \widehat{G}_{h_1}(\bx(\lambda),\bx_1,w_1)\widehat{G}_{h_2}(\bx(\lambda), \bx_2, w_2) \cA^{n-1}_{h_{12|n}h_3...h_n,\tilde{h}_2...\tilde{h}_m}(\bx(\lambda),\bx_3,...,\bx_n)
\vspace{3mm}
\\
\dps
+ \pi\Bigg(\int_{D_1}d^2\bx\;\sqrt{g(\bx)}\;\widetilde{G}_{h_1}(\bx_1,\bx)G_{h_2}(\bx_2,\bx) \cA^{n-1}_{\tilde{h}_1h_3...h_n,\tilde{h}_2...\tilde{h}_m}(\bx,\bx_3,...,\bx_n) + (1 \leftrightarrow 2)\Bigg)\,,
\ea
\ee 
where the domain $D_i\subset\RR\times\CC$ is the same integration domain as in the first term on the right-hand side of the superposition identity \eqref{superposition_id}
\be 
\label{D_def}
D_i = [0,u_i]\times[z_i-i(u-u_i),z_i+i(u-u_i)]\,;
\ee
the coefficients $a$, $\gamma$ and $\dm$ are given by \eqref{a_beta}, the $\beta$-coefficients are defined in \eqref{beta_coef} and the regularization factor $(1-\varepsilon)^n$ appears in the second term due to the interchange of the integral over $\bx$ in \eqref{general_struct} with the integral over $\lambda$ and the sum over $n$. The appearance of the regularization factor and the derivation of the coefficient  $\gamma_{\tilde{h}_1h_1h_2}$ using the following identity  \cite{Alkalaev:2026gkq}
\be 
\label{abel_sum}
\lim_{y\to1^-} \sum_{n=0}^\infty \frac{a^{h_1h_2}_n}{\dm^{h_3}_{h_{12|n}}}\frac{\beta_{h_3h_1h_2}}{\beta_{h_{12|n}h_1h_2}}y^n
= \gamma_{h_1h_2h_3}\,,
\ee
are part of the regularization procedure, see \cite{Alkalaev:2026gkq} for details. Note that the first weight of the $(n-1)$-point AdS Feynman diagram in the second term of \eqref{reduction_ex} differs from the first weight of the diagram in the other terms due to the following identity
\be 
\label{transition_diag}
\ba{l}
\dps
\int_{\text{AdS}_2}d^2\bx\;\sqrt{g(\bx)}\;G_h(\bx_1,\bx)\cA^{n-1}_{h_1h_2...h_{n-1},\tilde{h}}(\bx,\bx_2,...,\bx_{n-1}) = \frac{1}{\dm^{h_1}_{h}}\cA^{n-1}_{h_1h_2...h_{n-1},\tilde{h}}(\bx_1,\bx_2,...,\bx_{n-1})
\vspace{3mm}
\\
\dps
\hspace{81mm}+ \frac{1}{\dm^{h}_{h_1}}\cA^{n-1}_{hh_2...h_{n-1},\tilde{h}}(\bx_1,\bx_2,...,\bx_{n-1})\,,
\ea
\ee 
which directly follows from the transition identity \eqref{transition_id} and the representation of an $(n-1)$-point diagram as in \eqref{general_struct}. This identity is stated for real points $\bx_i\in\RR_{\geq0}\times\RR$, for $\bx_i\in\CC\times\CC$ one deforms the integration contours in \eqref{transition_diag} as in the transition identity \eqref{transition_id}. This completes the case $k=2$.

In the case $k \geq 3$, the application of the decomposition algorithm is more involved compared with the  case $k=2$. One starts with applying the superposition identity \eqref{superposition_id} to the bulk-to-bulk propagators $G_{h_1}(\bx_1,\bx)$ and $G_{h_2}(\bx_2,\bx)$ in the expression \eqref{general_struct}, which produces three terms. Converting bulk-to-bulk propagators into the modified propagators $\widehat{G}$ and applying the geodesic decomposition identity \eqref{geodesic_prop} to the product of modified propagators $\widehat{G}_{h_1}(\bx,\bx_1,w_1) $$\times\widehat{G}_{h_2}(\bx, \bx_2,w_2)$ in the first term, one obtains
\be
\label{cont_int_0}
\ba{l}
\dps
\cA^{n,k}_{h,\tilde{h}}(\bx_i) =  \lim_{\varepsilon\to 0^+}\sum_{n=0}^\infty\frac{a^{h_1h_2}_n(1-\varepsilon)^n}{\beta_{h_{12|n}h_1h_2}}\prod_{j=1}^2\int_{z_j-iu_j}^{z_j+iu_j}dw_j \int_{\gamma_{12}}d\lambda\; \widehat{G}_{h_1}(\bx(\lambda),\bx_1,w_1)\widehat{G}_{h_2}(\bx(\lambda), \bx_2,w_2) 
\vspace{3mm}
\\
\dps
\times \cA^{n-1}_{h_{12|n}h_3...h_n,\tilde{h}}(\bx(\lambda),\bx_3,...,\bx_n) \,
+ \pi\int_{D_1}d^2\bx\;\sqrt{g(\bx)}\; \widetilde{G}_{h_1}(\bx_1,\bx)G_{h_2}(\bx_2,\bx)...G_{h_k}(\bx_k,\bx)
\vspace{3mm}
\\
\dps
\times\cA^{n-k+1}_{\tilde{h}_1h_{k+1}...h_n,\tilde{h}_2...\tilde{h}_m}(\bx,\bx_{k+1},...,\bx_n) + (1 \leftrightarrow 2)\,,
\ea
\ee
where the factor $(1-\varepsilon)^n$ appears as the result of the regularization procedure, required to interchange the integration order in $\lambda$ and $\bx$, see \cite{Alkalaev:2026gkq} for details. We also define the $(n-1)$-point AdS Feynman diagram
\be 
\label{reduction_cont}
\ba{l}
\dps
\cA^{n-1}_{h_1...h_{n-1},\tilde{h}}(\bx_1,...,\bx_{n-1}) 
\vspace{3mm}
\\
\dps
\hspace{1mm}= \int_{\text{AdS}_2}d^2\bx\;\sqrt{g(\bx)}\; G_{h_1}(\bx_1,\bx)G_{h_2}(\bx_2,\bx)...G_{h_{k-1}}(\bx_{k-1},\bx)\cA^{n-k+1}_{\tilde{h}_1h_k...h_{n-1},\tilde{h}_2...\tilde{h}_m}(\bx,\bx_k,...,\bx_{n-1})\,.
\ea
\ee 
This is the main difference between the cases $k=2$ and $k\geq3$. In the case $k=2$, the original expression \eqref{general_struct} already has the form of the $(n-1)$-point AdS Feynman diagram integrated against two bulk-to-bulk propagators. The application of the first part of the decomposition algorithm preserves this structure, resulting in a sum of $(n-1)$-point AdS Feynman diagrams integrated against modified propagators \eqref{reduction_ex}. This is not the case for $k\geq3$, where applying the same steps results in a sum of the integrals of the $(n-1)$-point diagram (first term in \eqref{cont_int_0}) and the $(n-k+1)$-point diagram (second and third terms in \eqref{cont_int_0}). To express the result only in terms of $(n-1)$-point diagrams, one applies the conversion identity \eqref{conversion_id} to the bulk-to-bulk propagators $G_{h_3}(\bx_3,\bx)$, $G_{h_4}(\bx_4,\bx)$ and then uses the geodesic decomposition identity \eqref{geodesic_prop}, yielding
$$
\ba{l}
\dps
\cA^{n,k}_{h,\tilde{h}}(\bx_i) =  \lim_{\varepsilon\to 0^+}\sum_{n=0}^\infty\frac{a^{h_1h_2}_n(1-\varepsilon)^n}{\beta_{h_{12|n}h_1h_2}}\prod_{j=1}^2\int_{z_j-iu_j}^{z_j+iu_j}dw_j \int_{\gamma_{12}}d\lambda\; \widehat{G}_{h_1}(\bx(\lambda),\bx_1,w_1)\widehat{G}_{h_2}(\bx(\lambda), \bx_2,w_2) 
\ea 
$$
\be
\label{cont_int}
\ba{l}
\dps
\times \cA^{n-1}_{h_{12|n}h_3...h_n,\tilde{h}}(\bx(\lambda),\bx_3,...,\bx_n) 
+ \lim_{\varepsilon\to 0^+}\Bigg(\sum_{n=0}^\infty \frac{a^{h_3h_4}_n(1-\varepsilon)^n}{\beta_{h_{34|n}h_3h_4}} \prod_{j=3}^4\int_{z_j-iu_j}^{z_j+iu_j}dw_j \int_{\gamma_{34}}d\lambda\;
\vspace{3mm}
\\
\dps
\times \widehat{G}_{h_3}(\bx(\lambda),\bx_3,w_3)\widehat{G}_{h_4}(\bx(\lambda), \bx_4,w_4) \widetilde{\cA}^{n-1}_{h_1h_2h_{34|n}h_5...h_n,\tilde{h}}(\bx_1,\bx_2,\bx(\lambda),\bx_5,...,\bx_n) + (1 \leftrightarrow 2)\Bigg)\,,
\ea
\ee
where we have defined the modified $(n-1)$-point AdS Feynman diagram
\be 
\label{reduction_cont_mod}
\ba{l}
\dps
\widetilde{\cA}^{n-1}_{h_1...h_{n-1},\tilde{h}}(\bx_1,...,\bx_{n-1})
\vspace{3mm}
\\
\dps
\hspace{1mm}= \pi\int_{D_1}d^2\bx\;\sqrt{g(\bx)}\;\widetilde{G}_{h_1}(\bx_1,\bx) G_{h_2}(\bx_2,\bx)...G_{h_{k-1}}(\bx_{k-1},\bx)\cA^{n-k+1}_{\tilde{h}_1h_k...h_{n-1},\tilde{h}_2...\tilde{h}_m}(\bx,\bx_k,...,\bx_{n-1})\,.
\ea
\ee 
Note that in the case $k=3$, one applies the conversion and the geodesic decomposition identities to the propagators $G_{h_2}(\bx_2, \bx)$, $G_{h_3}(\bx_3, \bx)$ in the second term of \eqref{cont_int_0} and to $G_{h_1}(\bx_1, \bx)$, $G_{h_3}(\bx_3, \bx)$ in the third term, since the bulk-to-bulk propagator $G_{h_4}(\bx_4, \bx)$ is not present for $k=3$. 

Step I avoids the problem of a large number of intermediate terms by applying the superposition identity to a limited number of bulk-to-bulk propagators. As can be seen from \eqref{reduction_ex} and \eqref{cont_int}, the resulting expression has at most four terms regardless of the number of points in the original diagram.

\paragraph{II.} Step I, described in the previous paragraph, produces a sum of integrals involving $(n-1)$-point (modified) AdS Feynman diagrams. Step II consists of substitution of the Wilson network decomposition of these diagrams. If such a decomposition is unknown, one employs the reduction technique again to decompose these diagrams into AdS vertex functions. Since the reduction technique reduces the problem of decomposing an $n$-point diagram to that of an $(n-1)$-point diagram, in a finite number of applications of the reduction technique one can reduce the former problem to the problem of decomposing four-point diagrams, which has already been solved.

To demonstrate the application of step II, we consider the case $k=2$ \eqref{reduction_ex} and suppose that the Wilson network decomposition of the $(n-1)$-point diagram in \eqref{reduction_ex} takes the following form
\be 
\label{exp_n-1}
\cA^{n-1}_{\tilde{h}_1h_3...h_n,\tilde{h}_2...\tilde{h}_m}(\bx,\bx_3,...,\bx_n) = \sum_{i=1}^{N}\kappa_{i}\cV_{{\bf h_i},{\bf \tilde{h}_i}}(\bx,\bx_3,...,\bx_n)\,,
\ee 
where $N$ denotes the number of terms in the expansion, ${\bf h_i}$ and ${\bf \tilde{h}_i}$ denote a particular arrangement of the external and intermediate weights of the AdS vertex function in the $i$-th term. For brevity, we omit the infinite summations over the double-trace weights that appear in the decomposition. Substitution of this expansion into \eqref{reduction_ex} results in
$$
\ba{l}
\dps
\cA^{n,2}_{h,\tilde{h}}(\bx_i) = \frac{\gamma_{\tilde{h}_1h_1h_2}}{\beta_{\tilde{h}_1h_1h_2}}\sum_{i=1}^{N}\kappa_{i}\,\prod_{j=1}^2\int_{z_j-iu_j}^{z_j+iu_j}dw_j \int_{\gamma_{12}}d\lambda\; \widehat{G}_{h_1}(\bx(\lambda),\bx_1,w_1)\widehat{G}_{h_2}(\bx(\lambda), \bx_2,w_2)
\ea
$$
\be
\label{reduction_ex_int}
\ba{l}
\dps
\times\cV_{{\bf h_i},{\bf \tilde{h}_i}}(\bx(\lambda),\bx_3,...,\bx_n) + \lim_{\varepsilon\to 0^+}\sum_{n=0}^\infty\frac{a^{h_1h_2}_n(1-\varepsilon)^n}{\beta_{h_{12|n}h_1h_2}\dm^{h_{12|n}}_{\tilde{h}_1}}\sum_{i=1}^{N}\kappa_{i}\,\prod_{j=1}^2\int_{z_j-iu_j}^{z_j+iu_j}dw_j 
\vspace{3mm}
\\
\dps
\times \int_{\gamma_{12}}d\lambda\; \widehat{G}_{h_1}(\bx(\lambda),\bx_1,w_1)\widehat{G}_{h_2}(\bx(\lambda), \bx_2,w_2) \cV_{{\bf h_i'},{\bf \tilde{h}_i'}}(\bx(\lambda),\bx_3,...,\bx_n)
\vspace{3mm}
\\
\dps
+ \Bigg(\pi\sum_{i=1}^{N}\kappa_{i}\,\int_{D_1}d^2\bx\;\sqrt{g(\bx)}\;\widetilde{G}_{h_1}(\bx_1,\bx)G_{h_2}(\bx_2,\bx) \cV_{{\bf h_i},{\bf \tilde{h}_i}}(\bx,\bx_3,...,\bx_n) + (1 \leftrightarrow 2)\Bigg)\,,
\ea
\ee 
where we have introduced the notation ${\bf h_i'}$ and ${\bf \tilde{h}_i'}$ to account for the difference between the weights of the $(n-1)$-point AdS vertex functions in the second term and those in the other terms. 

Step II in the case $k\geq 3$ is carried out similarly, except that in addition to the Wilson network decomposition of the unmodified $(n-1)$-point AdS Feynman diagram one must also decompose the modified $(n-1)$-point diagram present in \eqref{reduction_cont}. Such a decomposition can be obtained by applying the reduction technique to the unmodified $(n-1)$-point AdS Feynman diagram and isolating the required modified diagram after step I.

\paragraph{III.} Step II produced a sum of integrals involving $(n-1)$-point AdS vertex functions. Step III converts these into $n$-point AdS vertex functions using the integral relations between $n$-point and $(n-1)$-point AdS vertex functions. While the exact form of these relations depends on the original $n$-point diagram, the application of step II always produces two types of terms: (a) integrals involving the product of the $(n-1)$-point AdS vertex function and a pair of modified propagators $\widehat{G}$, see the first two terms in \eqref{reduction_ex_int}; (b)  integrals involving the product of the $(n-1)$-point AdS vertex function and the propagators $\widetilde{G}G$, see the last two terms in \eqref{reduction_ex_int}. 

Thus, the problem of finding representations of the modified AdS Feynman diagrams that appeared in the previous decomposition method is replaced with the easier problem of finding two types of integral relations between AdS vertex functions. In the comb channel, the problem of finding these relations is further simplified, since the AdS vertex functions in the comb channel are reconstructed from the global conformal blocks \cite{Alkalaev:2024cje} which are structurally simpler than their bulk counterparts. One first finds relations between conformal blocks on the boundary and then uses the HKLL reconstruction to obtain the bulk relations between the AdS vertex functions. Using this method one can demonstrate that the integrals of type (a) are proportional to the $n$-point AdS vertex function
\be 
\label{n_geodesic}
\ba{l}
\dps
\cV_{h_1...h_n,\tilde{h}_1...\tilde{h}_{n-3}}(\bx_1,...,\bx_n) =\frac{1}{\beta_{\tilde{h}_1h_1h_2}}\prod_{j=1}^2\int_{z_j-iu_j}^{z_j+iu_j}dw_j\int_{\gamma_{12}} d\lambda 
\vspace{3mm}
\\
\dps
\hspace{25mm}\times \widehat{G}_{h_1}(\bx(\lambda),\bx_1,w_1)\widehat{G}_{h_2}(\bx(\lambda),\bx_2,w_2)\cV_{\tilde{h}_1h_3...h_n,\tilde{h}_2...\tilde{h}_{n-3}}(\bx(\lambda),\bx_3,...,\bx_n) \,,
\ea
\ee 
where each AdS vertex function is taken in the comb channel and the weights satisfy the following set of triangle inequalities:
\be 
h_1<\tilde{h}_1+h_2,\, h_2<\tilde{h}_1+h_1\,. 
\ee  
For the integrals of type (b), the method described above yields
\be 
\label{n_tilde}
\ba{l}
\dps
\sum_{n=0}^{\infty} \frac{a^{h_2\tilde{h}_1}_n}{\dm^{h_{2\tilde{1}|n}}_{h_1}}\cV_{h_{2\tilde{1}|n}h_2...h_n,\tilde{h}_1...\tilde{h}_{n-3}}(\bx_1,...,\bx_n) 
\vspace{3mm}
\\
\dps
=\pi\int_{D_1}d^2\bx\;\sqrt{g(\bx)}\; \widetilde{G}_{h_1}(\bx,\bx_1)G_{h_2}(\bx,\bx_2)\cV_{\tilde{h}_1h_3...h_n,\tilde{h}_2...\tilde{h}_{n-3}}(\bx,\bx_3,...,\bx_n) \,,
\ea
\ee 
where $h_{2\tilde{1}|n} = h_2+\tilde{h}_1+2n$ and each AdS vertex function is taken in the comb channel. The proofs of both relations are given in Appendix \bref{app:derivations}. 

These relations can be used to demonstrate the application of step III. Consider the $n$-point AdS Feynman diagram in the comb channel. This diagram belongs to the case $k=2$, and the application of steps I and II results in \eqref{reduction_ex_int}. The relation \eqref{n_geodesic} is applied to the first and second terms in \eqref{reduction_ex_int} and the relation \eqref{n_tilde} is applied to the last two terms in \eqref{reduction_ex_int}. The result is given by
\be
\label{reduction_comb}
\ba{l}
\dps
\cA^{n, \text{comb}}_{h,\tilde{h}}(\bx_i) = \frac{\gamma_{\tilde{h}_1h_1h_2}}{\beta_{\tilde{h}_1h_1h_2}} \sum_{i=1}^{N}\kappa_{i}\,\beta_{{\bf h_i}^1h_1h_2}\,\cV_{h_1h_2{\bf h_i}^{\cancel{1}},{\bf h_i}^1{\bf \tilde{h}_i}}(\bx_i)
\vspace{3mm}
\\
\dps
\hspace{21mm}+ \sum_{i=1}^{N}\lim_{\varepsilon\to 0^+}\sum_{n=0}^\infty\kappa_{i}\frac{a^{h_1h_2}_n(1-\varepsilon)^n}{\beta_{h_{12|n}h_1h_2}\dm^{h_{12|n}}_{\tilde{h}_1}}\,\beta_{{\bf h_i'}^1h_1h_2}\,\cV_{h_1h_2{\bf h_i'}^{\cancel{1}},{\bf h_i'}^1{\bf \tilde{h}_i'}}(\bx_i)
\vspace{3mm}
\\
\dps
\hspace{21mm}+ \Bigg(\sum_{i=1}^{N}\sum_{n=0}^{\infty} \kappa_{i}\frac{a^{h_2{\bf h_i}^1}_n}{\dm^{h_{2\,1_i|n}}_{h_1}}\,\cV_{h_{2\,1_i|n}h_2{\bf h_i}^{\cancel{1}},{\bf h_i}^1{\bf \tilde{h}_i}}(\bx_i) + (1 \leftrightarrow 2)\Bigg)\,,
\ea
\ee 
where ${\bf h_i}^1$ and ${\bf h_i'}^1$ are the first elements of the sets ${\bf h_i}$ and ${\bf h_i'}$, respectively, ${\bf h_i}^{\cancel{1}}$ and ${\bf h_i'}^{\cancel{1}}$ are obtained from ${\bf h_i}$ and ${\bf h_i'}$ by removing the first element, and $h_{2\,1_i|n}  = h_1 + {\bf h_i}^1 + 2n$. Since the decomposition of the $n$-point AdS Feynman diagram in the comb channel involves only that of the $(n-1)$-point diagram in the same channel, the relations \eqref{n_geodesic} and \eqref{n_tilde} constitute a complete set of integral relations between AdS vertex functions necessary for decomposing such $n$-point AdS Feynman diagram with arbitrary $n$. However, we do not provide a closed-form expression of the resulting decomposition due to its size and restrict our analysis to five-point AdS Feynman diagrams.

\section{Decompositions of five-point Feynman diagrams}
\label{sec:decomp}

\begin{figure}
\centering
\includegraphics[scale=0.9]{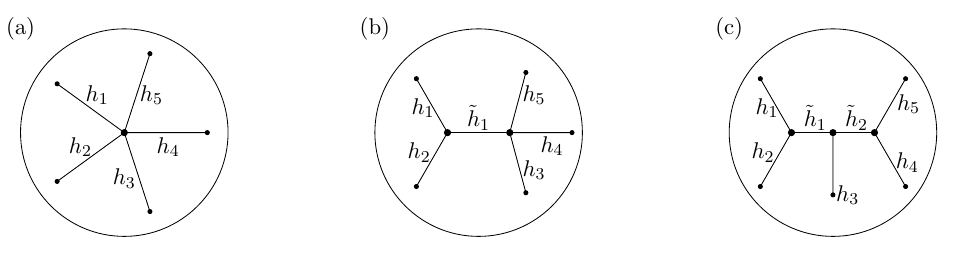}
\caption{All topologically distinct five-point AdS Feynman tree diagrams: {\bf (a)} contact diagram; {\bf (b)} single-exchange diagram; {\bf (c)} double-exchange diagram.}
\label{fig:5pt_diag}
\end{figure}
In this section, we apply the reduction technique to the three types of five-point AdS Feynman diagrams: {\bf (a)} contact,
\be 
\label{cont_def}
\dc{h}(\bx_i) = \int_{\text{AdS}_2}d^2\bx\;\sqrt{g(\bx)}\; \prod_{i=1}^5 G_{h_i}(\bx,\bx_i)\,,
\ee 
where bulk-to-bulk propagators are given by \eqref{bulk-to-bulk}. The following triangle inequalities ensure the convergence of the integrals appearing in the application of the reduction technique
\be
\label{triangle_identity1}
h_1 + \dots - h_k + \dots + h_5 > 0\,, 
\qquad 
\forall k \in \{1, ...\,, 5\}\,;
\ee
{\bf (b)} single-exchange,
\be 
\label{1-exch_def}
\cA^{\text{1-exch}}_{h,\tilde{h}}(\bx_i) = \iint_{\text{AdS}_2}d^2\bx d^2\bx'\sqrt{g(\bx)g(\bx')}\; \prod_{i=1}^2 G_{h_i}(\bx,\bx_i) G_{\tilde{h}_1}(\bx,\bx')\prod_{j=3}^5 G_{h_j}(\bx',\bx_j)\,,
\ee 
where the weights are restricted by the following inequalities
\be 
\label{triangle_identity2}
\ba{lll}
\dps h_i+h_j - h_k > 0  \,, & \qquad i\neq j\neq k\neq i\,, & \qquad i,j,k \in \{1,2,\tilde{1}\}\,,
\vspace{3mm}
\\
\dps h_3+h_4+h_5+\tilde{h}_1 - 2h_n > 0 \,, & \qquad n\in \{3,4,5,\tilde{1}\}\,, &
\ea 
\ee
where the symbol $\tilde{1}$ in the index sets denotes the intermediate weight $h_{\tilde{1}} \equiv \tilde h_1$; and {\bf (c)} double-exchange,
\be 
\label{2-exch_def}
\ba{l}
\dps
\cA^{\text{2-exch}}_{h,\tilde{h}}(\bx_i) =  \iiint_{\text{AdS}_2}d^2\bx d^2\bx'd^2\bx''\sqrt{g(\bx)g(\bx')g(\bx'')}
\vspace{3mm}
\\
\dps
\hspace{27mm}\times \prod_{i=1}^2 G_{h_i}(\bx,\bx_i) G_{\tilde{h}_1}(\bx,\bx')G_{h_3}(\bx',\bx_3)G_{\tilde{h}_2}(\bx',\bx'')\prod_{j=4}^5 G_{h_j}(\bx'',\bx_j)\,,
\ea
\ee 
where the weights satisfy the following triangle inequalities
\be 
\label{triangle_identity3}
\ba{lll}
\dps h_{i_1}+h_{i_2} - h_{i_3} > 0  \,, & \qquad i_1\neq i_2\neq i_3\neq i_1\,, & \qquad i_1,i_2,i_3 \in \{1,2,\tilde{1}\}\,,
\vspace{3mm}
\\
\dps h_{j_1}+h_{j_2} - h_{j_3} > 0  \,, & \qquad j_1\neq j_2\neq j_3\neq j_1\,, & \qquad j_1,j_2,j_3 \in \{\tilde{1},3,\tilde{2}\}\,,
\vspace{3mm}
\\
\dps h_{k_1}+h_{k_2} - h_{k_3} > 0  \,, & \qquad k_1\neq k_2\neq k_3\neq k_1\,, & \qquad k_1,k_2,k_3 \in \{\tilde{2},4,5\}\,.
\ea 
\ee

\subsection{Five-point contact diagram}
\label{sec:5_cont}

\paragraph{I.} Step I of the reduction technique produces the following three-term decomposition of the five-point contact AdS Feynman diagram
\be 
\label{5pt_cont_int}
\ba{l}
\dps
\dc{h}(\bx_i) = \lim_{\varepsilon\to0^+}\sum_{n=0}^\infty \frac{a^{h_1h_2}_n(1-\varepsilon)^n}{\beta_{h_{12|n}h_1h_2}} \prod_{j=1}^2\int_{z_j-iu_j}^{z_j+iu_j}dw_j\int_{\gamma_{12}}d\lambda\; \prod_{i=1}^2 \widehat{G}_{h_i}(\bx(\lambda),\bx_i,w_i)
\vspace{3mm}
\\
\dps
\times \dc{h_{12|n}h_3h_4h_5}(\bx(\lambda),\bx_3,\bx_4,\bx_5) + \lim_{\varepsilon\to0^+}\Big( \sum_{n=0}^\infty \frac{a^{h_4h_5}_n(1-\varepsilon)^n}{\beta_{h_{45|n}h_4h_5}}\prod_{j=4}^5\int_{z_j-iu_j}^{z_j+iu_j}dw_j\int_{\gamma_{45}}d\lambda\; 
\vspace{3mm}
\\
\dps
\times \prod_{i=4}^5\widehat{G}_{h_i}(\bx(\lambda),\bx_i,w_i)\tdc{h_1h_2h_3h_{45|n}}(\bx_1,\bx_2,\bx_3,\bx(\lambda))  + (1 \leftrightarrow 2) \Big) \,,
\ea 
\ee 
where $\dc{h_{12|n}h_3h_4h_5}(\bx(\lambda),\bx_3,\bx_4,\bx_5)$ is the four-point contact AdS Feynman diagram defined in \eqref{4pt_cont_decomp} and $\tdc{h_1h_2h_3h_{45|n}}(\bx_1,\bx_2,\bx_3,\bx(\lambda))$ is a modified four-point diagram 
\be 
\tdc{h_1h_2h_3h_4}(\bx_1,...,\bx_4) = \pi\int_{D_1}d^2\bx\;\sqrt{g(\bx)}\;\widetilde{G}_{h_1}(\bx,\bx_1)\prod_{i=2}^4 G_{h_i}(\bx,\bx_i) \,,
\ee 
where the domain $D_1$ is given by \eqref{D_def}. The regularization factors $(1-\varepsilon)^n$ appeared due to the interchange of the order of sums and integrals, see details of the regularization procedure below equation \eqref{reduction_ex} and in \cite{Alkalaev:2026gkq}. Note that the decomposition \eqref{5pt_cont_int} can be directly obtained from the $n$-point decomposition \eqref{reduction_cont} by substituting $n=5$, taking into account the topology of the contact diagram and relabelling the indices of conformal weights and coordinates. 

\paragraph{II-III.} To apply step II, one uses the Wilson network decompositions of the four-point AdS Feynman diagrams appearing in \eqref{5pt_cont_int}. The decomposition of the four-point contact AdS Feynman diagram in the first term of \eqref{5pt_cont_int} is given by \eqref{4pt_cont_decomp}. For the modified diagram in the second term, the corresponding decomposition reads
\be 
\label{modified_cont_dec}
\tdc{h_1h_2h_3h_4}(\bx_1,...,\bx_4)  = \sum_{n,m=0}^\infty\frac{a^{h_3h_4}_ma^{h_{34|m}h_2}_n}{\dm^{h_{234|nm}}_{h_1}}\cV_{h_{234|nm}h_2h_3h_4,h_{34|m}}(\bx_1,...,\bx_4)\,,
\ee 
which follows from the decomposition of the four-point AdS Feynman diagram \cite{Alkalaev:2026gkq}. Substituting the decompositions \eqref{modified_cont_dec} and \eqref{4pt_cont_decomp} into \eqref{5pt_cont_int} yields a sum of geodesic integrals of the four-point AdS vertex functions. Note that the weights of the four-point AdS Feynman diagram in the first term of \eqref{5pt_cont_int} do not necessarily satisfy the triangle inequalities \eqref{triangle_identity_4pt}. Nevertheless, we use the expansion \eqref{4pt_cont_decomp} since it can be analytically continued to the relevant conformal weights. Step III converts these integrals into the five-point AdS vertex functions using \eqref{n_geodesic}. The result of the application of these steps to \eqref{5pt_cont_int} is given by
\be 
\label{cont_final}
\ba{l}
\dps
\dc{h}(\bx_i)= \sum_{n,m=0}^{\infty} a^{h_1h_2}_n \gamma_{h_{12|n}h_3h_{45|m}}a^{h_4h_5}_m\, 
\cV_{h,h_{12|n}h_{45|m}}(\bx_i)
\vspace{3mm}
\\
\dps
\hspace{20mm}
+\sum_{n,m=0}^{\infty} a^{h_1h_2}_n a^{h_{12|n}h_3}_m \gamma_{h_{123|nm}h_4h_5}\,
\cV_{h,h_{12|n}h_{123|nm}}(\bx_i)
\vspace{3mm}
\\
\dps
\hspace{20mm}
+
\sum_{n,m=0}^{\infty} \gamma_{h_1h_2h_{345|mn}} a^{h_{45|m}h_3}_n a^{h_4h_5}_m \, 
\cV_{h,h_{345|mn}h_{45|m}}(\bx_i)
\vspace{3mm}
\\
\dps
\hspace{20mm}
+\Big(\sum_{n,m,k=0}^{\infty}
\frac{a^{h_{345|mn}h_2}_ka^{h_{45|m}h_3}_n a^{h_4h_5}_m}{\dm^{h_{2345|nmk}}_{h_1}}\,
\cV_{h_{2345|nmk} h_2 h_3 h_4 h_5,h_{345|mn}h_{45|m}}(\bx_i) + (1\leftrightarrow 2)\Big)
\vspace{3mm}
\\
\dps
\hspace{20mm}
+\sum_{n,m,k=0}^{\infty}
\frac{a^{h_1h_2}_na^{h_{12|n}h_{45|m}}_k a^{h_4h_5}_m}{\dm^{h_{1245|mnk}}_{h_3}}\,
\cV_{h_1 h_2 h_{1245|mnk} h_4 h_5,h_{12|n}h_{45|m}}(\bx_i)
\vspace{3mm}
\\
\dps
\hspace{20mm}
+\Big(\sum_{n,m,k=0}^{\infty}
\frac{a^{h_1h_2}_na^{h_{12|n}h_3}_m a^{h_{123|nm}h_5}_k}{\dm^{h_{1235|nmk}}_{h_4}}\,
\cV_{h_1 h_2 h_3 h_{1235|nmk} h_5,h_{12|n}h_{123|nm}}(\bx_i) + (4\leftrightarrow 5)\Big)\,,
\ea
\ee 
where we have used \eqref{abel_sum} to simplify the coefficients. This completes the Wilson network decomposition of the five-point contact AdS Feynman diagram.

The decomposition \eqref{cont_final} has a standard structure, where the external weights of the AdS vertex functions in the first three terms are single-trace, i.e. those coinciding with the original external weights $h_i$, while the AdS vertex functions in the remaining terms have at least one "multi-trace" weight $h_{ijkl|mnk}$ as one of their external weights. Below, we demonstrate that in the boundary limit the multi-trace terms are sub-leading compared with the first three terms.

\subsection{Five-point exchange diagrams}
\label{sec:5_ex}

To apply the reduction technique to the five-point diagrams with at least one exchange, one writes them in the following form
\be 
\label{5_exch_interm}
\ba{l}
\dps
\cA^{\text{1-exch}}_{h,\tilde{h}}(\bx_i)= \int_{\text{AdS}_2} d^2\bx\;\sqrt{g(\bx)}\;G_{h_1}(\bx,\bx_1)G_{h_2}(\bx,\bx_2)\cA^{\text{cont}}_{\tilde{h}_1h_3h_4h_5}(\bx, \bx_3,\bx_4,\bx_5)\,,
\vspace{3mm}
\\
\dps
\cA^{\text{2-exch}}_{h,\tilde{h}}(\bx_i)= \int_{\text{AdS}_2} d^2\bx \;\sqrt{g(\bx)}\;G_{h_1}(\bx,\bx_1)G_{h_2}(\bx,\bx_2)\cA^{\text{exch}}_{\tilde{h}_1h_3h_4h_5,\tilde{h}_2}(\bx, \bx_3,\bx_4,\bx_5)\,,
\ea
\ee 
where $\cA^{\text{cont/exch}}$ are the standard four-point AdS Feynman diagrams, either contact or exchange, depending on the number of exchanges (1 or 2) in the original diagram. Since the decompositions of the single-exchange and double-exchange diagrams have the same structure, we treat them simultaneously.

\paragraph{I.} Applying step I of the reduction technique to \eqref{1-exch_def} and \eqref{2-exch_def} yields
\be 
\label{5pt_interm_exp}
\ba{l}
\dps
\cA^{\text{1-/2-exch}}_{h,\tilde{h}}(\bx_i) = \frac{\gamma_{\tilde{h}_1h_1h_2}}{\beta_{\tilde{h}_1h_1h_2}} \prod_{j=1}^2\int_{z_j-iu_j}^{z_j+iu_j}dw_j \int_{\gamma_{12}}d\lambda\;\widehat{G}_{h_1}(\bx(\lambda),\bx_1,w_1)\widehat{G}_{h_2}(\bx(\lambda),\bx_2,w_2)
\vspace{3mm}
\\
\dps
\times \cA_{\tilde{h}_1h_3h_4h_5, \tilde{h}_2}^{\text{cont/exch}}(\bx(\lambda), \bx_3,\bx_4,\bx_5) + \lim_{\varepsilon\to0^+}\sum_{n=0}^\infty\frac{a^{h_1h_2}_n(1-\varepsilon)^n}{\beta_{h_{12|n}h_1h_2}}\frac{1}{\dm_{\tilde{h}_1}^{h_{12|n}}} \prod_{j=1}^2\int_{z_j-iu_j}^{z_j+iu_j}dw_j
\vspace{3mm}
\\
\dps
\times \int_{\gamma_{12}}d\lambda\; \widehat{G}_{h_1}(\bx(\lambda),\bx_1,w_1)\widehat{G}_{h_2}(\bx(\lambda),\bx_2,w_2) \cA_{h_{12|n}h_3h_4h_5, \tilde{h}_2}^{\text{cont/exch}}(\bx(\lambda), \bx_3,\bx_4,\bx_5)
\vspace{3mm}
\\
\dps
+  \pi\int_{D_1}d^2\bx\;\sqrt{g(\bx)}\;
\widetilde{G}_{h_1}(\bx,\bx_1) G_{h_2}(\bx,\bx_2)\cA_{\tilde{h}_1h_3h_4h_5, \tilde{h}_2}^{\text{cont/exch}}(\bx, \bx_3,\bx_4,\bx_5) + (1\leftrightarrow 2 )\,,
\ea
\ee 
where $D_1$ is given by \eqref{D_def}. This decomposition can be obtained from the $n$-point decomposition \eqref{reduction_ex} by setting $n=5$ and using the topology of the original five-point diagrams.

\paragraph{II-III.} Step II consists in the substitution of the Wilson network decompositions of the four-point contact \eqref{4pt_cont_decomp} or exchange (see \cite{Alkalaev:2026gkq}) diagrams into \eqref{5pt_interm_exp}. The result is a sum of integrals of four-point AdS vertex functions. To apply step III, one converts these integrals into five-point AdS vertex functions using the relations \eqref{n_geodesic} and \eqref{n_tilde}. The resulting decomposition of the five-point single-exchange AdS Feynman diagram is given by
$$
\ba{l}
\dps
\cA^{\text{1-exch}}_{h,\tilde{h}}(\bx_i) = \sum_{n=0}^\infty \kappa^{\text{1-ex}}_1(n)\cV_{h,\tilde{h}_1h_{45|n}} + \sum_{n=0}^\infty \kappa^{\text{1-ex}}_2(n)\cV_{h,\tilde{h}_1h_{\tilde{1}3|n}}+ \sum_{n,m=0}^\infty \kappa^{\text{1-ex}}_3(n,m)\cV_{h,h_{12|m}h_{45|n}}
\vspace{3mm}
\\
\dps
+\sum_{n,m=0}^\infty \kappa^{\text{1-ex}}_4(n,m)\cV_{h,h_{12|m}h_{123|nm}}
+\sum_{n,m=0}^\infty \kappa^{\text{1-ex}}_5(n,m)\cV_{h,h_{345|nm}h_{45|n}}
\vspace{3mm}
\\
\dps
+ \Bigg(\sum_{n,m=0}^\infty \kappa^{\text{1-ex}}_6(n,m)\cV_{h_{2\tilde{1}|m}h_2...h_5,\tilde{h}_1h_{45|n}}
+\sum_{n,m=0}^\infty \kappa^{\text{1-ex}}_{7}(n,m)\cV_{h_{2\tilde{1}|m}h_2...h_5,\tilde{h}_1h_{\tilde{1}3|n}}
\vspace{3mm}
\\
\dps
+\sum_{n,m,k=0}^\infty \kappa^{\text{1-ex}}_{8}(n,m,k)\cV_{h_{2\tilde{1}|m}h_2h_{\tilde{1}45|nk}h_4h_5,\tilde{h}_1h_{45|n}}
+\Big(\sum_{n,m,k=0}^\infty \kappa^{\text{1-ex}}_{9}(n,m,k)\cV_{h_{2\tilde{1}|m}h_2h_3h_{\tilde{1}35|nk}h_5,\tilde{h}_1h_{\tilde{1}3|n}}
\vspace{3mm}
\\
\dps
+(4 \leftrightarrow 5)\Big)
+\sum_{n,m,k=0}^\infty \kappa^{\text{1-ex}}_{11}(n,m,k)\cV_{h_{2345|nmk}h_2...h_5,h_{345|nm}h_{45|n}}+(1 \leftrightarrow 2)\Bigg)
\ea
$$
\be
\label{1-ex_final}
\ba{l}
\dps
+\Bigg(\sum_{n,m=0}^\infty \kappa^{\text{1-ex}}_{18}(n,m)\cV_{h_1h_2h_3h_{\tilde{1}35|nm}h_5,\tilde{h}_1h_{\tilde{1}3|n}}
+\sum_{n,m,k=0}^\infty \kappa^{\text{1-ex}}_{19}(n,m,k)\cV_{h_1...h_4h_{1234|nmk},h_{12|n}h_{123|nm}} +(4 \leftrightarrow 5)\Bigg)
\vspace{3mm}
\\
\dps
+\sum_{n,m=0}^\infty \kappa^{\text{1-ex}}_{22}(n,m)\cV_{h_1h_2h_{\tilde{1}45|nm}h_4h_5,\tilde{h}_1h_{45|n}}+\sum_{n,m,k=0}^\infty \kappa^{\text{1-ex}}_{23}(n,m,k)\cV_{h_1h_2h_{1245|nmk}h_4h_5,h_{12|n}h_{45|m}}\,,
\ea
\ee 
where we have used the identity \eqref{abel_sum} and the following identity
\be 
\frac{1}{\dm^{h_1}_{h_2}\dm^{h_3}_{h_1}} + \frac{1}{\dm^{h_2}_{h_1}\dm^{h_3}_{h_2}} = \frac{1}{\dm^{h_3}_{h_1}\dm^{h_3}_{h_2}}\,,
\ee 
to simplify the coefficients in the expansion. The coefficients of the expansion are given in Appendix \bref{app:coefs}. For brevity, we omit the $\bx$-dependence of the AdS vertex functions, since this dependence is the same for each term, i.e. $\cV_{h,h_{12|m}h_{123|nm}} = \cV_{h,h_{12|m}h_{123|nm}}(\bx_i)$. 

The Wilson network decomposition of the five-point double-exchange AdS Feynman diagram has $56$ terms, therefore, we write it in schematic form
$$
\ba{l}
\dps
\cA^{\text{2-exch}}_{h,\tilde{h}}(\bx_i)= \kappa^{\text{2-ex}}_1\cV_{h,\tilde{h}} + \sum_{n=0}^\infty \kappa^{\text{2-ex}}_2(n)\cV_{h,\tilde{h}_1h_{45|n}}+ \sum_{n=0}^\infty \kappa^{\text{2-ex}}_3(n)\cV_{h,h_{12|n}\tilde{h}_2} + \sum_{n=0}^\infty \kappa^{\text{2-ex}}_4(n)\cV_{h,h_{3\tilde{2}|n}\tilde{h}_2}
\vspace{3mm}
\\
\dps
+ \sum_{n=0}^\infty \kappa^{\text{2-ex}}_5(n)\cV_{h,\tilde{h}_1h_{\tilde{1}3|n}}
+ \sum_{n,m=0}^\infty \kappa^{\text{2-ex}}_6(n,m)\cV_{h,h_{12|n}h_{123|nm}}
+ \sum_{n,m=0}^\infty \kappa^{\text{2-ex}}_7(n,m)\cV_{h,h_{345|nm}h_{45|n}}
\ea
$$
\be
\label{2-ex_final}
\ba{l}
\dps
+ \sum_{n,m=0}^\infty \kappa^{\text{2-ex}}_8(n,m)\cV_{h,h_{12|n}h_{45|m}}
+ \sum_{s=1}^{5}\sum_{n=0}^\infty \kappa^{\text{2-ex, 1 d-t}}_s(n)\cV_{\sigma^{\text{1 d-t}}_s(h,\tilde{h})}
\vspace{3mm}
\\
\dps
+ \sum_{s=1}^{22}\sum_{n,m=0}^\infty \kappa^{\text{2-ex, 2 d-t}}_s(n,m)\cV_{\sigma^{\text{2 d-t}}_s(h,\tilde{h})}
+ \sum_{s=1}^{21}\sum_{n,m,k=0}^\infty \kappa^{\text{2-ex, 3 d-t}}_s(n,m,k)\cV_{\sigma^{\text{3 d-t}}_s(h,\tilde{h})}\,,
\ea
\ee 
where the coefficients and the indices $\sigma$ corresponding to the arrangements of the weights of the AdS vertex functions are given in Appendix \bref{app:coefs}. The first eight terms do not have double-trace external weights. We show that these terms are leading when calculating the boundary asymptotics of the AdS Feynman diagram. The rest of the terms are grouped by the number of double-trace weights in the AdS vertex function, as reflected in the notation of the coefficients: $1$ d-t (one double-trace weight) and so on. 

As discussed in part III of Section \bref{sec:reduction}, the reduction technique can be applied to decompose the $n$-point AdS Feynman diagram in the comb channel. The decomposition of such a diagram \eqref{reduction_comb} generalizes the decomposition of the five-point double-exchange diagram \eqref{2-ex_final}. Using the five-point case as a base case, one can recursively find the coefficients and the arrangements of the weights in \eqref{reduction_comb} for arbitrary $n$.

\subsection{Algorithm for constructing expansions of the AdS Feynman diagrams}
\label{sec:rules}

The resulting decompositions are rather extensive. Instead of writing down all terms of these expansions, one can employ an algorithm for constructing the expansion of the diagram from its graph. This algorithm is based on the analysis of the expansions of three-, four- and five-point diagrams, such as \eqref{4pt_cont_decomp}, \eqref{cont_final}, \eqref{1-ex_final}, \eqref{2-ex_final}.

\begin{figure}
\centering
\includegraphics[scale=0.9]{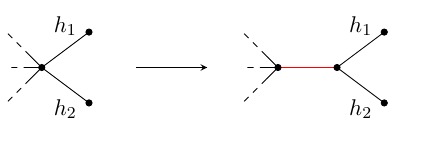}
\caption{The transformation that lowers the vertex degree by one. This operation corresponds to applying the geodesic decomposition identity \eqref{geodesic_prop}, which replaces two modified propagators by the geodesic integral of three propagators.}
\label{fig:geod_dec}
\end{figure}

\begin{figure}
\centering
\includegraphics[scale=0.9]{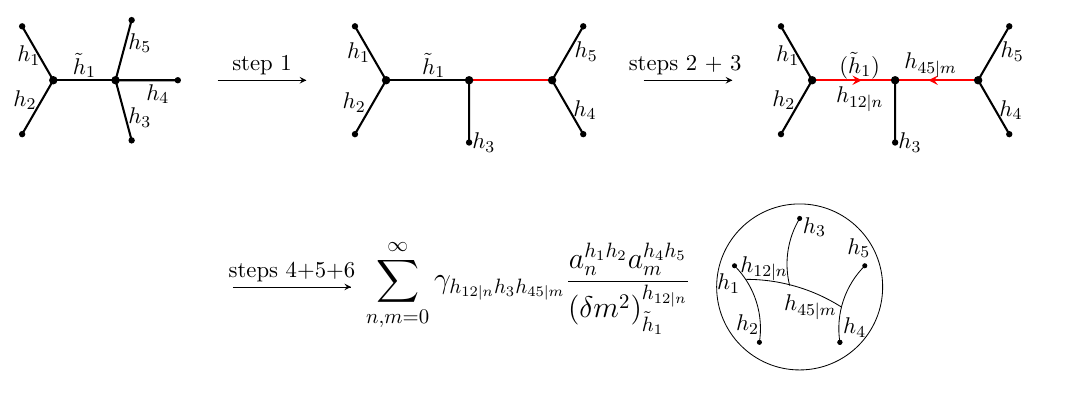}
\caption{Example of applying the algorithm to the five-point single-exchange AdS Feynman diagram. After the first step, the original diagram is converted into the simplified diagram consisting of only cubic vertices. The application of the second and the third steps produces the arrangement of the weights in the resulting AdS vertex function and allows one to fix the coefficient of the expansion. The resulting term in the expansion is obtained after applying steps 4, 5 and 6. The graph in the second line denotes the matrix element of Wilson line network, consisting of  Wilson lines (curved lines) connected by $\sltwo$ intertwiners. The weights are given by $h_{i...j|n...m} = h_i+...+h_j+2n+...+2m$.}
\label{fig:example}
\end{figure}

\begin{enumerate}
    \item Consider the $n$-point AdS Feynman diagram with $n\leq 5$. Each external line carries the conformal weight $h_i$ and each internal line carries the conformal weight $\tilde{h}_j$. If the diagram contains a non-cubic vertex, one applies the transformation shown in fig.~\bref{fig:geod_dec} to any two lines converging at such a vertex. The meaning of the red lines will be explained later. This operation lowers the valency of the vertex by one and produces a new cubic vertex. Repeat this step until no four-valent or five-valent vertices remain. The resulting graph, which we call the simplified diagram, corresponds to the graph of an AdS vertex function with weights to be determined.
    \item To obtain the coefficient and the arrangement of the weights of the AdS vertex function, one first colours the simplified diagram using the following rules:
    \begin{itemize}
        \item Each red line of the simplified diagram produced at step 1 must be replaced with a red line with an arrow. For each such red line there are two possible directions of this arrow. 
        \item Any black line can be replaced with a red one with an arrow. The conformal weight carried by the original line should not be discarded but instead put in parentheses. It will be used later when determining the coefficient.
        \item Each vertex can have at most one outgoing red line and any number of incoming red lines.
        \item Arrows on the external red lines must point outwards, towards the endpoint. 
    \end{itemize}
    Different colourings of the diagram and different choices of arrow directions correspond to different coefficients and arrangements of the weights of the resulting AdS vertex functions.
    \item The red lines carry the double-trace weights. To determine the weight of each red line, one starts with a vertex with one outgoing red line. The weight carried by this line is $h_{12|n}$, where $h_1$ and $h_2$ are the weights of the other two lines in this vertex and $n$ is a new summation variable that is not present in the other double-trace weights. If any of the lines with the weights $h_1$ or $h_2$ is red, then its weight may be undetermined. In such a case, one should first find the weight of this undetermined line and only then the weight of the original line. Repeating this procedure for every vertex with an outgoing red line, one fixes the weights of all red lines provided that the coloured simplified diagram is built according to the above rules. See fig.~\bref{fig:example} for an example of the application of the present and the previous steps.
    \item The arrangement of the weights of the AdS vertex function is read off from the weights carried by the lines of the coloured simplified diagram.
    \item There are only two elements of the coloured simplified diagram contributing to the coefficient of the AdS vertex function: vertices and red lines with weights in parentheses (i.e. black lines in the simplified diagram replaced with the red ones after colouring). Starting from the vertices, if a vertex in the diagram has an outgoing red line with the weight $h_{12|n}$, then the factor $a^{h_1h_2}_n$ is present in the coefficient, where $h_1$ and $h_2$ are the weights of the other two lines in the vertex. In any other case, a vertex contributes the factor $\gamma_{h_1h_2h_3}$, where $h_1$, $h_2$ and $h_3$ are the weights of the three lines converging at this vertex. The red lines with the weights in parentheses contribute an additional factor $\frac{1}{\dm^{h_1}_{h_2}}$ to the coefficient, where $h_1$ is the weight of the red line and $h_2$ is the weight in parentheses. The coefficient of the AdS vertex function corresponding to the diagram shown in fig.~\bref{fig:example} is given by
    \be 
    \kappa(n,m) = \gamma_{h_{12|n}h_3h_{45|m}}\frac{a^{h_1h_2}_n a^{h_4h_5}_m}{\dm^{h_{12|n}}_{\tilde{h}_1}}
    \ee 
    \item A term of the expansion produced by a particular coloured simplified diagram is given by a sum of AdS vertex functions associated with this diagram with the coefficient $\kappa(n_k)$ and summed over all $n_k$ appearing in the double-trace weights $h_{ij|n_k}$. The term produced by the diagram from fig.~\bref{fig:example} is:
    \be 
    \sum_{n,m=0}^\infty \kappa(n,m)\cV_{h_1h_2h_3h_4h_5,h_{12|n}h_{45|m}}(\bx_i)
    \ee 
    \item The sum of the terms obtained above over all possible colourings (including changes in the directions of the arrows) of the simplified diagram produces the complete expansion of the AdS Feynman diagram.
\end{enumerate}

\subsection{Conformal boundary asymptotics}
\label{sec:boundary}

To verify that our decomposition is consistent with the conformal block expansions of the Witten diagrams from \cite{Jepsen:2019svc}, we study the boundary asymptotics of the obtained expansions. Starting with the five-point contact diagram, one takes the limit $u\to0$ in \eqref{cont_final} and obtains
$$
\ba{l}
\dps
\lim_{u_1=...=u_5=u\to 0} u^{-h_1...-h_5}\dc{h_1...h_5}(\bx_i)= \sum_{n,m=0}^{\infty} a^{h_1h_2}_n \gamma_{h_{12|n}h_3h_{45|m}}a^{h_4h_5}_m\, 
F_{h,h_{12|n}h_{45|m}}(z_i)
\ea
$$
\be 
\label{bndry_5pt_cont}
\ba{l}
\dps
+\sum_{n,m=0}^{\infty} a^{h_1h_2}_n a^{h_{12|n}h_3}_m \gamma_{h_{123|nm}h_4h_5}\,
F_{h,h_{12|n}h_{123|nm}}(z_i)
\vspace{3mm}
\\
\dps
+
\sum_{n,m=0}^{\infty} \gamma_{h_1h_2h_{345|mn}} a^{h_{45|m}h_3}_n a^{h_4h_5}_m \, 
F_{h,h_{345|mn}h_{45|m}}(z_i)\,,
\ea
\ee
where $F_{h,\tilde{h}}(z_i)$ is the five-point global conformal block. To obtain \eqref{bndry_5pt_cont}, we have imposed the following triangle-type restrictions on the conformal weights:
\be 
h_1 + \dots - h_k + \dots + h_5 > 0\,, 
\qquad 
\forall k \in \{1, ...\,, 5\}\,.
\ee
The boundary asymptotics of the five-point AdS Feynman diagram \eqref{1-ex_final} is given by
$$
\ba{l}
\dps
\lim_{u_1=...=u_5=u\to 0} u^{-h_1...-h_5}\cA^{\text{1-exch}}_{h_1...h_5,\tilde{h}_1}(\bx_i)= \sum_{m=0}^{\infty} \gamma_{h_1h_2\tilde{h}_1} \gamma_{\tilde{h}_1h_3h_{45|m}}a^{h_4h_5}_m\, 
F_{h,\tilde{h}_1h_{45|m}}(z_i)
\ea 
$$
\be 
\label{bndry_5pt_1_ex}
\ba{l}
\dps
+\sum_{m=0}^{\infty} \gamma_{h_1h_2\tilde{h}_1} a^{h_3\tilde{h}_1}_m \gamma_{h_{3\tilde{1}|m}h_4h_5}\,
F_{h,\tilde{h}_1h_{3\tilde{1}|m}}(z_i)
\vspace{3mm}
\\
\dps
+
\sum_{n,m=0}^{\infty} \frac{a^{h_1h_2}_n}{\dm^{h_{12|n}}_{\tilde{h}_1}}\gamma_{h_{12|n}h_3h_{45|m}} a^{h_4h_5}_m \, 
F_{h,h_{12|n}h_{45|m}}(z_i)
\vspace{3mm}
\\
\dps
+
\sum_{n,m=0}^{\infty} \frac{a^{h_1h_2}_n}{\dm^{h_{12|n}}_{\tilde{h}_1}}a^{h_{12|n}h_3}_m\gamma_{h_{123|nm}h_4h_5}  \, 
F_{h,h_{12|n}h_{123|nm}}(z_i)
\vspace{3mm}
\\
\dps
+
\sum_{n,m=0}^{\infty} \frac{\gamma_{h_1h_2h_{345|mn}}}{\dm^{h_{345|mn}}_{\tilde{h}_1}}a^{h_{45|m}h_3}_n a^{h_4h_5}_m \, 
F_{h,h_{345|mn}h_{45|m}}(z_i)\,,
\ea
\ee
where 
\be 
\ba{ll}
\dps h_i+h_j - h_k > 0  \,, & \qquad i\neq j\neq k\neq i\,,  \qquad i,j,k \in \{1,2,\tilde{1}\}\,,
\vspace{3mm}
\\
\dps h_{\sigma(3)}+h_{\sigma(4)}+h_{\sigma(5)} - h_{\sigma(\tilde{1})} > 0 \,, & \qquad \forall \text{ permutations } \sigma \text{ of indices }(3,4,5,\tilde{1})\,,
\ea 
\ee
where the symbol  $\tilde{1}$ in the index sets denotes the intermediate weight $h_{\tilde{1}} \equiv \tilde h_1$. The boundary asymptotics of the five-point AdS Feynman diagram with two exchanges reads
$$
\ba{l}
\dps
\lim_{u_1=...=u_5=u\to 0} u^{-h_1...-h_5}\cA^{\text{2-exch}}_{h_1...h_5,\tilde{h}_1\tilde{h}_2}(\bx_i)
\vspace{3mm}
\\
\dps
\hspace{40mm}
=  \gamma_{h_1h_2\tilde{h}_1} \gamma_{\tilde{h}_1h_3\tilde{h}_2}\gamma_{\tilde{h}_2h_4h_5}\, 
F_{h,\tilde{h}_1\tilde{h}_2}(z_i)
\vspace{3mm}
\\
\dps
\hspace{40mm}
+\sum_{n=0}^{\infty} \gamma_{h_1h_2\tilde{h}_1} \gamma_{\tilde{h}_1h_3h_{45|n}} \frac{a^{h_4h_5}_n}{\dm^{h_{45|n}}_{\tilde{h}_2}}\,
F_{h,\tilde{h}_1h_{45|n}}(z_i)
\vspace{3mm}
\\
\dps
\hspace{40mm}
+
\sum_{n=0}^{\infty} \frac{a^{h_1h_2}_n}{\dm^{h_{12|n}}_{\tilde{h}_1}} \gamma_{h_{12|n}h_3\tilde{h}_2}\gamma_{\tilde{h}_2h_4h_5} \, 
F_{h,h_{12|n}\tilde{h}_2}(z_i)
\vspace{3mm}
\\
\dps
\hspace{40mm}
+
\sum_{n=0}^{\infty} \gamma_{h_1h_2\tilde{h}_1} \frac{a^{\tilde{h}_1h_3}_n}{\dm^{h_{3\tilde{1}|n}}_{\tilde{h}_2}}\gamma_{h_{3\tilde{1}|n}h_4h_5}  \, 
F_{h,\tilde{h}_1h_{3\tilde{1}|n}}(z_i)
\vspace{3mm}
\\
\dps
\hspace{40mm}
+\sum_{n=0}^{\infty}  \gamma_{h_1h_2h_{3\tilde{2}|n}} \frac{a^{h_3\tilde{h}_2}_n}{\dm^{h_{3\tilde{2}|n}}_{\tilde{h}_1}} \gamma_{\tilde{h}_2h_4h_5}\, 
F_{h,h_{3\tilde{2}|n}\tilde{h}_2}(z_i)
\vspace{3mm}
\\
\dps
\hspace{40mm}
+
\sum_{n,m=0}^{\infty} \frac{a^{h_1h_2}_n}{\dm^{h_{12|n}}_{\tilde{h}_1}} \frac{a^{h_3h_{12|n}}_m}{\dm^{h_{123|mn}}_{\tilde{h}_2}} \gamma_{h_{123|nm}h_4h_5}  \, 
F_{h,h_{12|n}h_{123|nm}}(z_i)
\vspace{3mm}
\\
\dps
\hspace{40mm}
+
\sum_{n,m=0}^{\infty}  \gamma_{h_1h_2h_{345|nm}} \frac{a^{h_{45|m}h_3}_n}{\dm^{h_{345|nm}}_{\tilde{h}_1}} \frac{a^{h_4h_5}_m}{\dm^{h_{45|m}}_{\tilde{h}_2}}\, 
F_{h,h_{345|nm}h_{45|m}}(z_i)
\ea 
$$
\be 
\label{bndry_5pt_2_ex}
\ba{l}
\dps
+\sum_{n,m=0}^{\infty}   \frac{a^{h_1h_2}_n}{\dm^{h_{12|n}}_{\tilde{h}_1}}\gamma_{h_{12|n}h_3h_{45|m}} \frac{a^{h_4h_5}_m}{\dm^{h_{45|m}}_{\tilde{h}_2}}\, 
F_{h,h_{12|n}h_{45|m}}(z_i)\,,
\ea
\ee
where 
\be 
\ba{lll}
\dps h_{i_1}+h_{i_2} - h_{i_3} > 0  \,, & \qquad i_1\neq i_2\neq i_3\neq i_1\,, & \qquad i_1,i_2,i_3 \in \{1,2,\tilde{1}\}\,,
\vspace{3mm}
\\
\dps h_{j_1}+h_{j_2} - h_{j_3} > 0 \,, & \qquad j_1\neq j_2\neq j_3\neq j_1\,, & \qquad j_1,j_2,j_3 \in \{\tilde{1},3,\tilde{2}\}\,,
\vspace{3mm}
\\
\dps h_{k_1}+h_{k_2} - h_{k_3} > 0 \,, & \qquad k_1\neq k_2\neq k_3\neq k_1\,, & \qquad k_1,k_2,k_3 \in \{\tilde{2},4,5\}\,.
\ea 
\ee
These three expansions coincide with the expansions of the corresponding five-point Witten diagrams given in \cite{Jepsen:2019svc}.\footnote{Note that the triangle inequalities in every case are necessary not only for consistency of the results but also for the convergence of the Witten diagrams. Otherwise, the regularisation of the integrals is necessary; see \cite{Castro:2024cmf} for details regarding the three-point Witten diagram.}

\section{Conclusion}
\label{sec:conclusion}

In this paper, we have proposed the reduction technique for obtaining the Wilson network expansion of $n$-point AdS Feynman diagrams in two dimensions. The technique circumvents the two main obstructions that arise in the $n$-point generalization of the previous decomposition algorithm: a large number of intermediate terms that grows rapidly with $n$ and the necessity of complex integral relations between modified AdS$_2$ Feynman diagrams and AdS$_2$ vertex functions, because the algorithm acts only on external propagators. 

We formulated two integral relations between the $n$-point AdS$_2$ vertex functions in the comb channel and $(n-1)$-point ones, which form a complete set of relations required for decomposing an $n$-point diagram in the comb channel. To demonstrate the reduction technique, we have considered all three topologically distinct five-point AdS$_2$ Feynman tree diagrams. In each case, we have obtained the Wilson network expansion in terms of AdS$_2$ vertex functions with single-trace and multi-trace weights. At the conformal boundary, the resulting expansions reproduce the known conformal block decompositions of the corresponding five-point Witten diagrams \cite{Jepsen:2019svc}. We have also formulated an algorithm for constructing individual terms of the expansion directly from the graph of a diagram.

Looking ahead, the application of the reduction technique to an arbitrary $n$-point diagram requires formulating the integral relations between the $(n-1)$-point and the $n$-point AdS$_2$ vertex functions. Ultimately, it is expected that these relations can be found for any topology of AdS$_2$ vertex functions and we leave this for future work.

Motivated by the $1/N$ expansion, we have restricted our analysis to tree diagrams. However, the reduction technique can be applied to loop diagrams that can be reduced to tree diagrams using the method of \cite{Fitzpatrick:2011hu}. This approach employs  a K{\"a}llen-Lehmann-type representation of powers of the bulk-to-bulk propagator, effectively replacing the loop with an infinite  sum of tree propagators. Extending the reduction technique to loop diagrams that cannot be simplified this way remains an interesting problem for future work. One promising avenue is to employ the Mellin representation, see e.g. \cite{Penedones:2010ue, Fitzpatrick:2011hu}. Although typically formulated only for Witten diagrams, the Mellin representation significantly simplifies their evaluation. We leave both the formulation of the Mellin representation for AdS Feynman diagrams with bulk points and the subsequent application of the reduction technique to future work.

\noindent \textbf{Acknowledgements.}  The author is grateful to Konstantin Alkalaev for useful discussions and to Evgenii Vasiliev for valuable comments on the manuscript. This work was supported by the Foundation for the Advancement of Theoretical Physics and Mathematics “BASIS”.

\appendix

\section{Derivation of integral relations for AdS vertex functions}
\label{app:derivations}

\paragraph{AdS vertex functions in the comb channel integrated with two modified propagators $\widehat{G}$.} To prove the relation \eqref{n_geodesic}, we first consider its boundary version:
\be 
\label{bndry_n_geodesic}
\ba{l}
\dps
\cV^{\text{reg}}_{h_1...h_n,\tilde{h}_1...\tilde{h}_{n-3}}(z_1,...,z_n) 
\vspace{3mm}
\\
\dps
= \frac{1}{\beta_{\tilde{h}_1h_1h_2}}\int_{\gamma_{12}} d\lambda \,\cV^{\text{reg}}_{\tilde{h}_1h_3...h_n,\tilde{h}_2...\tilde{h}_{n-3}}(\bx(\lambda),z_3,...,z_n) K_{h_1}(\bx(\lambda),z_1)K_{h_2}(\bx(\lambda),z_2)\,,
\ea
\ee 
where, for convenience, we have introduced the $n$-point regularized AdS vertex function with $k$ of its points on the boundary
\be
\label{regular}
\ba{l}
\dps
\cV_{ h,\tilde{h}}^{\text{reg}}(z_1,...,z_k,\bx_{k+1},...,\bx_{n})=\lim_{u\to 0}u^{-\sum_{i=1}^k h_i}\cV_{ h,\tilde{h}}(\bx_1,...,\bx_n)\big|_{u_1=...=u_k=u}\,.
\ea
\ee
Note that here and everywhere below we assume that the AdS vertex functions and conformal blocks are taken in the comb channel. The $n$-point regularized AdS vertex function with all points on the boundary coincides with the $n$-point global conformal block due to the extrapolate dictionary relation \cite{Alkalaev:2023axo}:
\be
\label{extrapol}
\lim_{u\to 0}u^{-\sum_{i=1}^nh_i}\,\cV_{h,\tilde{h}}(\bx_1,...,\bx_n)\Big|_{u_1=...=u_n = u}= \,F_{h, \tilde{h}}(z_1,...,z_n)\,.
\ee
By substituting the HKLL representation of the $(n-1)$-point regularized AdS vertex function \cite{Alkalaev:2024cje}
\be 
\label{HKLL_vertex_bndry}
\cV^{\text{reg}}_{\tilde{h}_1h_3...h_n,\tilde{h}_2...\tilde{h}_{n-3}}(\bx,z_3,...,z_n) = 
\int_{z-iu}^{z+iu}dw\;\mathbb{K}_{\tilde{h}_1}(\bx,w) F_{\tilde{h}_1h_3...h_n,\tilde{h}_2...\tilde{h}_{n-3}}(w,z_3,...,z_n)\,,
\ee 
into the right-hand side of \eqref{bndry_n_geodesic}, one obtains 
\be 
\label{interm_bndry_n_geodesic}
\ba{l}
\dps
\cV^{\text{reg}}_{h_1...h_n,\tilde{h}_1...\tilde{h}_{n-3}}(z_1,...,z_n) = \frac{1}{\beta_{\tilde{h}_1h_1h_2}}\int_{\gamma_{12}} d\lambda \,\int_{z(\lambda)-iu(\lambda)}^{z(\lambda)+iu(\lambda)}dw\;
\vspace{3mm}
\\
\dps
\times\mathbb{K}_{\tilde{h}_1}(\bx(\lambda),w) F_{\tilde{h}_1h_3...h_n,\tilde{h}_2...\tilde{h}_{n-3}}(w,z_3,...,z_n) K_{h_1}(\bx(\lambda),z_1)K_{h_2}(\bx(\lambda),z_2)\,,
\ea
\ee 
where $\mathbb{K}_{h}(\bx,w)$ is the smearing function \eqref{smear}.
The $w$-dependence of the conformal block can be factored out of the block using the following recursion relation \cite{Rosenhaus:2018zqn} 
\be 
\label{block_recursion}
\ba{l}
\dps
F_{\tilde{h}_1h_3...h_n,\tilde{h}_2...\tilde{h}_{n-3}}(w,z_3,...,z_n) = \sum_{k=0}^\infty\frac{(\tilde{h}_3+\tilde{h}_2-h_4)_k(\tilde{h}_2+\tilde{h}_1-h_3)_k}{k!(2\tilde{h}_2)_k}
\vspace{3mm}
\\
\dps
\times \left(\frac{z_{34}(w-z_3)}{(w-z_4)}\right)^{\tilde{h}_2+\tilde{h}_1-h_3+k}(w-z_3)^{-2\tilde{h}_1}F_{\tilde{h}_2+k\,h_4...h_n,\tilde{h}_3...\tilde{h}_{n-3}}(z_3,...,z_n) \,,
\ea
\ee 
where $z_{ij} = z_i-z_j$. To integrate over $u$, one substitutes \eqref{block_recursion} into \eqref{interm_bndry_n_geodesic} and uses the HKLL representation of the regularized three-point AdS vertex function
\be 
\label{HKLL_vertex_bndry_3pt}
\cV^{\text{reg}}_{h_1\, h_2\, h_3}(z_1,z_2,\bx)= z_{12}^{-2h_2}\int_{z-iu}^{z+iu}dw\;\mathbb{K}_{h_3}(\bx,w) \left(\frac{z_{12}(w-z_1)}{(w-z_2)}\right)^{h_2+h_3-h_1}(w-z_1)^{-2h_3}\,.
\ee 
The result is given by
\be 
\label{interm_bndry_n_geodesic_2}
\ba{l}
\dps
\cV^{\text{reg}}_{h_1...h_n,\tilde{h}_1...\tilde{h}_{n-3}}(z_1,...,z_n) 
\vspace{3mm}
\\
\dps
= \frac{1}{\beta_{\tilde{h}_1h_1h_2}}\sum_{k=0}^\infty\int_{\gamma_{12}} d\lambda \,\cV^{\text{reg}}_{h_3\, \tilde{h}_2+k\, \tilde{h}_1}(z_3,z_4,\bx(\lambda))K_{h_1}(\bx(\lambda),z_1)K_{h_2}(\bx(\lambda),z_2)
\vspace{3mm}
\\
\dps
\times \frac{(\tilde{h}_3+\tilde{h}_2-h_4)_k(\tilde{h}_2+\tilde{h}_1-h_3)_k}{k!(2\tilde{h}_2)_k} z_{34}^{2\tilde{h}_2+2k}\; F_{\tilde{h}_2+k\,h_4...h_n,\tilde{h}_3...\tilde{h}_{n-3}}(z_3,...,z_n)\,,
\ea
\ee 
To complete the proof, one uses the geodesic representation of the four-point conformal block \cite{Alkalaev:2026gkq}
\be 
\label{geodesic_rep_4pt}
F_{h_1 h_2 h_3 h_4,h}(z_1,z_2,z_3,z_4) = \frac{1}{\beta_{h h_3 h_4}} \int_{\gamma_{34}}d\lambda\, \cV^{\text{reg}}_{h_1 h_2 h}(z_1,z_2, \bx(\lambda))\,K_{h_3}(\bx(\lambda),z_3)\,K_{h_4}(\bx(\lambda),z_4)\,,
\ee  
where the conformal weights satisfy the triangle inequalities 
\be 
h_3<h+h_4,\quad h_4<h+h_3;
\ee 
along with another recursion relation for the conformal block \cite{Rosenhaus:2018zqn}
\be
\label{block_recursion_alt}
\ba{l}
\dps
F_{h_1...h_n,\tilde{h}_1...\tilde{h}_{n-3}}(z_1,...,z_n) = \sum_{k=0}^\infty \frac{(\tilde{h}_3+\tilde{h}_2-h_4)_k(\tilde{h}_2+\tilde{h}_1-h_3)_k}{k!(2\tilde{h}_2)_k}z_{34}^{2\tilde{h}_2+2k}
\vspace{3mm}
\\
\dps
\hspace{43mm}\times F_{h_1h_2h_3\tilde{h}_2+k,\tilde{h}_1}(z_1,...,z_4) F_{\tilde{h}_2+k\,h_4...h_n,\tilde{h}_3...\tilde{h}_{n-3}}(z_3,...,z_n)\,,
\ea
\ee 
and substitutes the resulting relation into \eqref{interm_bndry_n_geodesic_2}, yielding the $n$-point global conformal block
\be 
\ba{l}
\dps
\cV^{\text{reg}}_{h_1...h_n,\tilde{h}_1...\tilde{h}_{n-3}}(z_1,...,z_n)  = F_{h_1...h_n,\tilde{h}_1...\tilde{h}_{n-3}}(z_1,...,z_n)\,,
\ea
\ee 
which proves \eqref{bndry_n_geodesic} due to the extrapolate dictionary relation \eqref{extrapol}. The bulk version \eqref{n_geodesic} of the identity \eqref{bndry_n_geodesic} can be obtained by multiplying both sides of \eqref{bndry_n_geodesic} by the smearing functions $\mathbb{K}_{h_i}$ for $i\in\{1,\dots,n\}$ and integrating over the complex boundary points $z_i\in\CC$.

\paragraph{AdS vertex function in the comb channel integrated against modified propagators $\widetilde{G}G$.} The proof of the relation \eqref{n_tilde} is analogous to the proof of \eqref{n_geodesic}. One starts by proving the boundary version of the identity \eqref{n_tilde} in which one point is kept in the bulk
\be 
\label{n_tilde_bndry}
\ba{l}
\dps
\sum_{n=0}^{\infty} \frac{a^{h_2\tilde{h}_1}_n}{\dm^{h_{2\tilde{1}|n}}_{h_1}}\cV^{\text{reg}}_{h_{2\tilde{1}|n}h_2...h_n,\tilde{h}_1...\tilde{h}_{n-3}}(\bx_1,z_2,...,z_n) 
\vspace{3mm}
\\
\dps
=\pi\int_{D_1}d^2\bx\;\sqrt{g(\bx)}\; \widetilde{G}_{h_1}(\bx,\bx_1)K_{h_2}(\bx,z_2)\cV^{\text{reg}}_{\tilde{h}_1h_3...h_n,\tilde{h}_2...\tilde{h}_{n-3}}(\bx,z_3,...,z_n) \,,
\ea
\ee 
where the point $\bx_1$ is located in the bulk, since the modified propagator $\widetilde{G}$ cannot be reconstructed from the boundary using the HKLL reconstruction. Despite this, one can apply the same steps as in the previous proof to \eqref{n_tilde_bndry} and obtain an analogue of \eqref{interm_bndry_n_geodesic_2}
\be 
\label{interm_bndry_n_tilde}
\ba{l}
\dps
\sum_{n=0}^{\infty} \frac{a^{h_2\tilde{h}_1}_n}{\dm^{h_{2\tilde{1}|n}}_{h_1}}\cV^{\text{reg}}_{h_{2\tilde{1}|n}h_2...h_n,\tilde{h}_1...\tilde{h}_{n-3}}(\bx_1,z_2,...,z_n)
\vspace{3mm}
\\
\dps
= \pi\sum_{k=0}^\infty\int_{D_1}d^2\bx\;\sqrt{g(\bx)}\;\widetilde{G}_{h_1}(\bx,\bx_1)K_{h_2}(\bx,z_2) \cV^{\text{reg}}_{\tilde{h}_1\,h_3\, \tilde{h}_2+k}(\bx, z_3,z_4)
\vspace{3mm}
\\
\dps
\times \frac{(\tilde{h}_3+\tilde{h}_2-h_4)_k(\tilde{h}_2+\tilde{h}_1-h_3)_k}{k!(2\tilde{h}_2)_k} z_{34}^{2\tilde{h}_2+2k}\; F_{\tilde{h}_2+k\,h_4...h_n,\tilde{h}_3...\tilde{h}_{n-3}}(z_3,...,z_n)\,.
\ea
\ee 
To integrate over $\bx$ one uses the following identity
\be 
\label{poch_rep_4pt}
\ba{l}
\dps
\sum_{n=0}^{\infty} \frac{a^{h_2\tilde{h}_1}_n}{\dm^{h_{2\tilde{1}|n}}_{h_1}} \cV^{\text{reg}}_{h_{2\tilde{1}|n} h_2 h_3 h_4,\tilde{h}_1}(\bx_1,z_2,z_3,z_4)
\vspace{3mm}
\\
\dps
\hspace{55mm}= \pi\int_{D_1}d^2\bx\;\sqrt{g(\bx)}\;\widetilde{G}_{h_1}(\bx,\bx_1)K_{h_2}(\bx,z_2) \cV^{\text{reg}}_{\tilde{h}_1h_3 h_4}(\bx, z_3,z_4)\,,
\ea
\ee 
which directly follows from the decomposition of the four-point contact AdS Feynman diagram \cite{Alkalaev:2026gkq}. Using the HKLL representation of the four-point AdS vertex function \eqref{HKLL_vertex_bndry} and the recursion relation for the conformal block \eqref{block_recursion_alt}, one obtains \eqref{n_tilde_bndry}

The proof of \eqref{n_tilde} now consists of multiplying both sides of \eqref{n_tilde_bndry} by $n-1$ smearing functions $\mathbb{K}_{h_i}$ for $i\in\{2,\dots,n\}$ and integrating the product over the complex boundary points $z_i\in\CC$.

\section{List of coefficients of the expansions }
\label{app:coefs}

\subsection{Coefficients for \eqref{1-ex_final}}

The coefficients in the expansion \eqref{1-ex_final} are given by
$$
\ba{ll}
\dps \kappa^{\text{1-ex}}_1(n) =  \gamma_{h_1h_2\tilde{h}_1}\gamma_{\tilde{h}_1h_3h_{45|n}}a^{h_4h_5}_n \,, &\quad \dps  \kappa^{\text{1-ex}}_2(n) =  \gamma_{h_1h_2\tilde{h}_1}a^{\tilde{h}_1h_3}_n\gamma_{h_{\tilde{1}3|n}h_4h_5} \,,
\ea 
$$
\newpage
\be 
\ba{ll}
\dps \kappa^{\text{1-ex}}_3(n,m) =  \frac{a^{h_1h_2}_m}{\dm^{h_{12|m}}_{\tilde{h}_1}}\gamma_{h_{12|m}h_3h_{45|n}}a^{h_4h_5}_n \,, &\quad \dps  \kappa^{\text{1-ex}}_4(n,m) =  \frac{a^{h_1h_2}_m}{\dm^{h_{12|m}}_{\tilde{h}_1}}a^{h_{12|m}h_3}_n \gamma_{h_{123|nm}h_4h_5} \,,
\vspace{3mm}
\\
\dps \kappa^{\text{1-ex}}_5(n,m) =   \gamma_{h_1h_2h_{345|nm}}\frac{a^{h_3h_{45|n}}_m}{\dm^{h_{345|nm}}_{\tilde{h}_1}}a^{h_4h_5}_n  \,, &\quad \dps  \kappa^{\text{1-ex}}_6(n,m) =  a^{h_2\tilde{h}_1}_m \gamma_{\tilde{h}_1h_3h_{45|n}} a^{h_4h_5}_n \,,
\vspace{3mm}
\\
\dps \kappa^{\text{1-ex}}_7(n,m) =  a^{h_2\tilde{h}_1}_m a^{\tilde{h}_1h_3}_n\gamma_{h_{\tilde{1}3|n}h_4h_5}   \,, &\quad \dps  \kappa^{\text{1-ex}}_8(n,m,k) =  a^{h_2\tilde{h}_1}_m a^{\tilde{h}_1h_{45|n}}_k a^{h_4h_5}_n \,,
\vspace{3mm}
\\
\dps \kappa^{\text{1-ex}}_9(n,m,k) =  a^{h_2\tilde{h}_1}_m a^{\tilde{h}_1h_3}_n a^{h_{\tilde{1}3}h_5}_k   \,, &\quad \dps  \kappa^{\text{1-ex}}_{10}(n,m,k) =  \kappa^{\text{1-ex}}_9(n,m,k)\big|_{h_4\leftrightarrow h_5} \,,
\vspace{3mm}
\\
\dps \kappa^{\text{1-ex}}_{11}(n,m,k) =  \frac{a^{h_2h_{345|nm}}_k}{\dm^{h_{345|nm}}_{\tilde{h}_1}} a^{h_3h_{45|n}}_m a^{h_4h_5}_n   \,, &\quad \dps  \kappa^{\text{1-ex}}_{i}(n,m,k) =  \kappa^{\text{1-ex}}_{i-6}(n,m,k)\big|_{h_1\leftrightarrow h_2}\,, \quad i \in \{12,...\,,17\} \,,
\vspace{3mm}
\\
\dps \kappa^{\text{1-ex}}_{18}(n,m) =   \gamma_{h_1h_2\tilde{h}_1} a^{\tilde{h}_1h_3}_n a^{h_{\tilde{1}3|n}h_5}_m    \,, &\quad \dps  \kappa^{\text{1-ex}}_{19}(n,m,k) = \frac{a^{h_1h_2}_n}{\dm^{h_{12|n}}_{\tilde{h}_1}}a^{h_{12|n}h_3}_ma^{h_{123|nm}h_4}_k\,,
\vspace{3mm}
\\
\dps \kappa^{\text{1-ex}}_{20}(n,m) =   \kappa^{\text{1-ex}}_{18}(n,m)\big|_{h_4\leftrightarrow h_5}   \,, &\quad \dps  \kappa^{\text{1-ex}}_{21}(n,m,k) =   \kappa^{\text{1-ex}}_{19}(n,m,k)\big|_{h_4\leftrightarrow h_5} \,,
\vspace{3mm}
\\
\dps \kappa^{\text{1-ex}}_{22}(n,m) =   \gamma_{h_1h_2\tilde{h}_1} a^{\tilde{h}_1h_{45|n}}_m a^{h_4h_5}_n   \,, &\quad \dps  \kappa^{\text{1-ex}}_{23}(n,m,k) = \frac{a^{h_1h_2}_n}{\dm^{h_{12|n}}_{\tilde{h}_1}}a^{h_{12|n}h_{45|m}}_k a^{h_4h_5}_m\,.
\ea 
\ee
These coefficients are in agreement with the algorithm formulated at the end of Section \bref{sec:rules}.

\subsection{Coefficients and weight arrangements for \eqref{2-ex_final}}

The first eight coefficients in \eqref{2-ex_final} are given by
\be 
\ba{ll}
\dps \kappa^{\text{2-ex}}_1=  \gamma_{h_1h_2\tilde{h}_1}\gamma_{\tilde{h}_1h_3\tilde{h}_2}\gamma_{\tilde{h}_2h_4h_5} \,, &\quad \dps  \kappa^{\text{2-ex}}_2(n) =  \gamma_{h_1h_2\tilde{h}_1}\gamma_{\tilde{h}_1h_3\tilde{h}_2}\frac{a^{h_4h_5}_n}{\dm^{h_{45|n}}_{\tilde{h}_2}} \,,
\vspace{3mm}
\\
\dps \kappa^{\text{2-ex}}_3(n) =  \frac{a^{h_1h_2}_n}{\dm^{h_{12|n}}_{\tilde{h}_1}}\gamma_{h_{12|n}h_3\tilde{h}_2}\gamma_{\tilde{h}_2h_4h_5} \,, &\quad \dps  \kappa^{\text{2-ex}}_4(n) =  \gamma_{h_1h_2h_{3\tilde{2}|n}}\frac{a^{h_3\tilde{h}_2}_n }{\dm^{h_{3\tilde{2}|n}}_{\tilde{h}_1}}\gamma_{\tilde{h}_2h_4h_5} \,,
\vspace{3mm}
\\
\dps \kappa^{\text{2-ex}}_5(n) =   \frac{\gamma_{h_1h_2\tilde{h}_1}a^{\tilde{h}_1h_3}_n }{\dm^{h_{\tilde{1}3|n}}_{\tilde{h}_2}}\gamma_{h_{\tilde{1}3|n}h_4h_5} \,, &\quad \dps  \kappa^{\text{2-ex}}_6(n,m) =  \frac{a^{h_1h_2}_n}{\dm^{h_{12|n}}_{\tilde{h}_1}}\frac{a^{h_{12|n}h_3}_m}{\dm^{h_{123|nm}}_{\tilde{h}_2}} \gamma_{h_{123|nm}h_4h_5} \,,
\vspace{3mm}
\\
\dps \kappa^{\text{2-ex}}_7(n,m) =  \frac{\gamma_{h_1h_2h_{345|nm}}a^{h_{45|n}h_3}_ma^{h_4h_5}_n}{\dm^{h_{345|nm}}_{\tilde{h}_1}\dm^{h_{45|n}}_{\tilde{h}_2}}   \,, &\quad \dps  \kappa^{\text{2-ex}}_8(n,m) =  \frac{a^{h_1h_2}_n}{\dm^{h_{12|n}}_{\tilde{h}_1}}  \gamma_{h_{12|n}h_3h_{45|m}} \frac{a^{h_4h_5}_m}{\dm^{h_{45|m}}_{\tilde{h}_2}} \,,
\ea 
\ee
The rest of the coefficients are provided with the corresponding AdS vertex function. The coefficients of the AdS vertex functions with only one double-trace conformal weight are given by
\be 
\ba{ll}
\dps \kappa^{\text{2-ex, 1 d-t}}_1(n) =  a^{\tilde{h}_1h_2}_n\gamma_{\tilde{h}_1h_3\tilde{h}_2}\gamma_{\tilde{h}_2h_4h_5}   \,, &\quad \dps  \cV_{\sigma^{\text{1 d-t}}_1(h,\tilde{h})} =  \cV_{h_{2\tilde{1}|n}h_2h_3h_4h_5,\tilde{h}_1\tilde{h}_2} \,,
\vspace{3mm}
\\
\dps \kappa^{\text{2-ex, 1 d-t}}_2(n) =  \kappa^{\text{2-ex, 1 d-t}}_1(n)\big|_{h_1\leftrightarrow h_2}   \,, &\quad \dps  \cV_{\sigma^{\text{1 d-t}}_2(h,\tilde{h})} =  \cV_{h_1h_{1\tilde{1}|n}h_3h_4h_5,\tilde{h}_1\tilde{h}_2} \,,
\vspace{3mm}
\\
\dps \kappa^{\text{2-ex, 1 d-t}}_3(n) =  \gamma_{h_1h_2\tilde{h}_1}a^{\tilde{h}_1\tilde{h}_2}_n\gamma_{\tilde{h}_2h_4h_5}    \,, &\quad \dps  \cV_{\sigma^{\text{1 d-t}}_3(h,\tilde{h})} =  \cV_{h_1h_2h_{\tilde{1}\tilde{2}|n}h_4h_5,\tilde{h}_1\tilde{h}_2} \,,
\vspace{3mm}
\\
\dps \kappa^{\text{2-ex, 1 d-t}}_4(n) =  \gamma_{h_1h_2\tilde{h}_1}\gamma_{\tilde{h}_1h_3\tilde{h}_2}a^{\tilde{h}_2h_5}_n  \,, &\quad \dps  \cV_{\sigma^{\text{1 d-t}}_4(h,\tilde{h})} =  \cV_{h_1h_2h_3h_{\tilde{2}5|n}h_5,\tilde{h}_1\tilde{h}_2} \,,
\vspace{3mm}
\\
\dps \kappa^{\text{2-ex, 1 d-t}}_5(n) = \kappa^{\text{2-ex, 1 d-t}}_4(n)\big|_{h_4\leftrightarrow h_5}    \,, &\quad \dps  \cV_{\sigma^{\text{1 d-t}}_5(h,\tilde{h})} =  \cV_{h_1h_2h_3h_4h_{\tilde{2}4|n},\tilde{h}_1\tilde{h}_2} \,.
\ea 
\ee
The coefficients of the AdS vertex functions with two double-trace weights are given by
\be 
\ba{ll}
\dps \kappa^{\text{2-ex, 2 d-t}}_1(n,m) =  a^{h_2\tilde{h}_1}_n\gamma_{\tilde{h}_1h_3\tilde{h}_2}a^{\tilde{h}_2h_5}_m   \,, &\quad \dps
\cV_{\sigma^{\text{2 d-t}}_1(h,\tilde{h})} =  \cV_{h_{2\tilde{1}|n}h_2h_3h_{\tilde{2}5|m}h_5,\tilde{h}_1\tilde{h}_2} \,,
\vspace{3mm}
\\
\dps \kappa^{\text{2-ex, 2 d-t}}_2(n,m) =  a^{h_1\tilde{h}_1}_n\gamma_{\tilde{h}_1h_3\tilde{h}_2}a^{\tilde{h}_2h_4}_m  \,, &\quad \dps
\cV_{\sigma^{\text{2 d-t}}_2(h,\tilde{h})} =  
\cV_{h_{1\tilde{1}|n}h_2h_3h_4h_{\tilde{2}4|m},\tilde{h}_1\tilde{h}_2} \,,
\vspace{3mm}
\\
\dps  \kappa^{\text{2-ex, 2 d-t}}_3(n,m) =  a^{h_2h_{3\tilde{2}|m}}_n\frac{a^{h_3\tilde{h}_2}_m }{\dm^{h_{3\tilde{2}|m}}_{\tilde{h}_1}}\gamma_{\tilde{h}_2h_4h_5}   \,, &\quad \dps  \cV_{\sigma^{\text{2 d-t}}_3(h,\tilde{h})} =  \cV_{h_{23\tilde{2}|nm}h_2h_3h_4h_5,h_{3\tilde{2}|m}\tilde{h}_2} \,,
\vspace{3mm}
\\
\dps \kappa^{\text{2-ex, 2 d-t}}_4(n,m) = a^{h_2\tilde{h}_1}_n\frac{a^{\tilde{h}_1h_3}_m }{\dm^{h_{\tilde{1}3|m}}_{\tilde{h}_2}}\gamma_{h_{\tilde{1}3|m}h_4h_5}    \,, &\quad \dps  \cV_{\sigma^{\text{2 d-t}}_4(h,\tilde{h})} =  \cV_{h_{2\tilde{1}|n}h_2h_3h_4h_5,\tilde{h}_1h_{\tilde{1}3|m}} \,,
\vspace{3mm}
\\
\dps \kappa^{\text{2-ex, 2 d-t}}_5(n,m) = a^{h_2\tilde{h}_1}_n\gamma_{\tilde{h}_1h_3h_{45|m}}\frac{a^{h_4h_5}_m }{\dm^{h_{45|m}}_{\tilde{h}_2}}  \,, &\quad \dps  \cV_{\sigma^{\text{2 d-t}}_5(h,\tilde{h})} =  \cV_{h_{2\tilde{1}|n}h_2h_3h_4h_5,\tilde{h}_1h_{45|m}} \,,
\vspace{3mm}
\\
\dps \kappa^{\text{2-ex, 2 d-t}}_6(n,m) = a^{h_2\tilde{h}_1}_n a^{\tilde{h}_1\tilde{h}_2}_m\gamma_{\tilde{h}_2h_4h_5}   \,, &\quad \dps  \cV_{\sigma^{\text{2 d-t}}_6(h,\tilde{h})} =  \cV_{h_{2\tilde{1}|n}h_2h_{\tilde{1}\tilde{2}|m}h_4h_5,\tilde{h}_1\tilde{h}_2} \,,
\vspace{3mm}
\\
\dps \kappa^{\text{2-ex, 2 d-t}}_7(n,m) = \frac{a^{h_1h_2}_n}{\dm^{h_{12|n}}_{\tilde{h}_1}}a^{h_{12|n}\tilde{h}_2}_m\gamma_{\tilde{h}_2h_4h_5}\,, &\quad \dps  \cV_{\sigma^{\text{2 d-t}}_7(h,\tilde{h})} =  \cV_{h_1h_2h_{12\tilde{2}|nm}h_4h_5,h_{12|n}\tilde{h}_2} \,,
\vspace{3mm}
\\
\dps \kappa^{\text{2-ex, 2 d-t}}_8(n,m) = \gamma_{h_1h_2\tilde{h}_1}a^{\tilde{h}_1h_{45|m}}_n\frac{a^{h_4h_5}_m}{\dm^{h_{45|m}}_{\tilde{h}_2}}    \,, &\quad \dps  \cV_{\sigma^{\text{2 d-t}}_8(h,\tilde{h})} =  \cV_{h_1h_2h_{\tilde{1}45|nm}h_4h_5,\tilde{h}_1h_{45|m}} \,,
\ea 
\ee
$$
\ba{l}
\dps 
\kappa^{\text{2-ex, 2 d-t}}_i(n,m) =  \kappa^{\text{2-ex, 2 d-t}}_{i-8}(n,m)\big|_{h_1\leftrightarrow h_2}  \,, 
\vspace{3mm}
\\
\dps \cV_{\sigma^{\text{2 d-t}}_i(h,\tilde{h})} =  \cV_{\sigma^{\text{2 d-t}}_{i-8}(h,\tilde{h})}\big|_{1\leftrightarrow 2} \,, \quad i\in\{9,...\,, 14\} \,,
\ea 
$$
\be 
\ba{l}
\dps
\kappa^{\text{2-ex, 2 d-t}}_j(n,m) =  \kappa^{\text{2-ex, 2 d-t}}_{j-12}(n,m)\big|_{h_1\leftrightarrow h_4, h_2\leftrightarrow h_5, \tilde{h}_1\leftrightarrow \tilde{h}_2}  \,, 
\vspace{3mm}
\\
\dps
\cV_{\sigma^{\text{2 d-t}}_j(h,\tilde{h})} =  \cV_{\sigma^{\text{2 d-t}}_{j-12}(h,\tilde{h})}\big|_{1\leftrightarrow 4, 2\leftrightarrow 5, \tilde{h}_1\leftrightarrow \tilde{h}_2} \,, \quad j\in\{15,...\,, 18\}\,,
\vspace{5mm}
\\
\dps \kappa^{\text{2-ex, 2 d-t}}_l(n,m) =  \kappa^{\text{2-ex, 2 d-t}}_{l-8}(n,m)\big|_{h_1\leftrightarrow h_4, h_2\leftrightarrow h_5, \tilde{h}_1\leftrightarrow \tilde{h}_2}  \,, 
\vspace{3mm}
\\
\dps
\cV_{\sigma^{\text{2 d-t}}_l(h,\tilde{h})} =  \cV_{\sigma^{\text{2 d-t}}_{l-8}(h,\tilde{h})}\big|_{1\leftrightarrow 4, 2\leftrightarrow 5, \tilde{h}_1\leftrightarrow \tilde{h}_2} \,, \quad l\in\{19,...\,, 22\}  \,,
\ea 
\ee
where the notation $1\leftrightarrow 2$ indicates the change of the indices of the conformal weights and coordinates in the AdS vertex function $\cV_{\sigma^{\text{2 d-t}}_s(h,\tilde{h})}$.

The coefficients of the AdS vertex functions with three double-trace weights are given by
\be 
\ba{ll}
\dps \kappa^{\text{2-ex, 3 d-t}}_1(n,m,k) =  a^{h_2\tilde{h}_1}_na^{\tilde{h}_1\tilde{h}_2}_ma^{\tilde{h}_2h_5}_k   \,, &\quad \dps
\cV_{\sigma^{\text{3 d-t}}_1(h,\tilde{h})} =  \cV_{h_{2\tilde{1}|n}h_2h_{\tilde{1}\tilde{2}|m}h_{\tilde{2}5|k}h_5,\tilde{h}_1\tilde{h}_2} \,,
\vspace{3mm}
\\
\dps \kappa^{\text{2-ex, 3 d-t}}_2(n,m,k)  =  a^{h_2h_{3\tilde{2}|m}}_n\frac{a^{h_3\tilde{h}_2}_m}{\dm^{h_{3\tilde{2}|m}}_{\tilde{h}_1}}a^{\tilde{h}_2h_5}_k   \,, &\quad \dps
\cV_{\sigma^{\text{3 d-t}}_2(h,\tilde{h})} =  
\cV_{h_{23\tilde{2}|nm}h_2h_3h_{\tilde{2}5|k}h_5,h_{3\tilde{2}|m}\tilde{h}_2} \,,
\vspace{3mm}
\\
\dps  \kappa^{\text{2-ex, 3 d-t}}_3(n,m,k)  =  a^{h_2\tilde{h}_1}_n\frac{a^{\tilde{h}_1h_3}_m}{\dm^{h_{\tilde{1}3|m}}_{\tilde{h}_2}}a^{h_{\tilde{1}3|m}h_5}_k   \,, &\quad \dps  \cV_{\sigma^{\text{3 d-t}}_3(h,\tilde{h})} =  \cV_{h_{2\tilde{1}|n}h_2h_3h_{\tilde{1}35|mk}h_5,\tilde{h}_1h_{\tilde{1}3|m}}  \,,
\vspace{3mm}
\\
\dps 
\kappa^{\text{2-ex, 3 d-t}}_i(n,m,k) =  \kappa^{\text{2-ex, 3 d-t}}_{i-3}(n,m)\big|_{h_4\leftrightarrow h_5}
\,, &\quad 
\cV_{\sigma^{\text{3 d-t}}_i(h,\tilde{h})} =  \cV_{\sigma^{\text{3 d-t}}_{i-3}(h,\tilde{h})}\big|_{4\leftrightarrow 5} \,, \quad i\in\{4, 5, 6\}
\vspace{3mm}
\\
\dps 
\kappa^{\text{2-ex, 3 d-t}}_7(n,m,k)  = \frac{a^{h_2h_{345|nm}}_ka^{h_3h_{45|n}}_ma^{h_4h_5}_n}{\dm^{h_{345|nm}}_{\tilde{h}_1}\dm^{h_{45|n}}_{\tilde{h}_2}}   
\,, &\quad \dps  \cV_{\sigma^{\text{3 d-t}}_7(h,\tilde{h})} =  \cV_{h_{2345|nmk}h_2h_3h_4h_5,h_{345|nm}h_{45|n}} \,,
\vspace{3mm}
\\
\dps \kappa^{\text{2-ex, 3 d-t}}_8(n,m,k) = a^{h_2\tilde{h}_1}_k a^{\tilde{h}_1h_{45|n}}_m\frac{a^{h_4h_5}_n}{\dm^{h_{45|n}}_{\tilde{h}_2}}\,, &\quad \dps  \cV_{\sigma^{\text{3 d-t}}_8(h,\tilde{h})} =   \cV_{h_{2\tilde{1}|k}h_2h_{\tilde{1}45|nm}h_4h_5,\tilde{h}_1h_{45|n}}\,,
\vspace{3mm}
\\
\dps 
\kappa^{\text{2-ex, 3 d-t}}_j(n,m,k) =  \kappa^{\text{2-ex, 3 d-t}}_{j-8}(n,m)\big|_{h_1\leftrightarrow h_2} \,, &\quad   \cV_{\sigma^{\text{3 d-t}}_j(h,\tilde{h})} =  \cV_{\sigma^{\text{3 d-t}}_{j-8}(h,\tilde{h})}\big|_{1\leftrightarrow 2} \,, \quad j\in\{9,...\,, 16\} \,,
\vspace{3mm}
\\
\dps \kappa^{\text{2-ex, 3 d-t}}_{17}(n,m,k) = \frac{a^{h_1h_2}_na^{h_{12|n}a_{45|m}}_ka^{h_4h_5}_m}{\dm^{h_{12|n}}_{\tilde{h}_1}\dm^{h_{45|m}}_{\tilde{h}_2}}   \,, &\quad \dps  \cV_{\sigma^{\text{3 d-t}}_{17}(h,\tilde{h})} =  \cV_{h_1h_2h_{1245|nmk}h_4h_5,h_{12|n}h_{45|m}} \,,
\ea 
\ee
\be 
\ba{l}
\dps 
\kappa^{\text{2-ex, 3 d-t}}_l(n,m,k) =  \kappa^{\text{2-ex, 3 d-t}}_{l-11}(n,m)\big|_{h_1\leftrightarrow h_4, h_2\leftrightarrow h_5, \tilde{h}_1\leftrightarrow \tilde{h}_2}  \,, 
\vspace{3mm}
\\
\dps
\cV_{\sigma^{\text{3 d-t}}_l(h,\tilde{h})} =  \cV_{\sigma^{\text{3 d-t}}_{l-11}(h,\tilde{h})}\big|_{1\leftrightarrow 4, 2\leftrightarrow 5, \tilde{h}_1\leftrightarrow \tilde{h}_2} \,, \quad l\in\{18, 19\}\,,
\vspace{5mm}
\\
\dps
\kappa^{\text{2-ex, 3 d-t}}_s(n,m,k) =  \kappa^{\text{2-ex, 3 d-t}}_{s-5}(n,m)\big|_{h_1\leftrightarrow h_4, h_2\leftrightarrow h_5, \tilde{h}_1\leftrightarrow \tilde{h}_2}  \,, 
\vspace{3mm}
\\
\dps
\cV_{\sigma^{\text{3 d-t}}_s(h,\tilde{h})} =  \cV_{\sigma^{\text{3 d-t}}_{s-5}(h,\tilde{h})}\big|_{1\leftrightarrow 4, 2\leftrightarrow 5, \tilde{h}_1\leftrightarrow \tilde{h}_2} \,, \quad s\in\{20, 21\}\,.
\ea 
\ee

\bibliographystyle{JHEP}
\bibliography{refs}

\end{document}